\documentclass[aps,prl,twocolumn,superscriptaddress]{revtex4}
\usepackage{graphicx}
\graphicspath{{./figures}}
\usepackage{color}
\usepackage{mhchem}

\begin{document}

\title{Tracking the surface atomic motion in a coherent phonon oscillation}

\author{ Davide Curcio}
\thanks{These two authors contributed equally}
\affiliation{Department of Physics and Astronomy, Interdisciplinary Nanoscience Center (iNANO), Aarhus University, 8000 Aarhus C, Denmark}

\author{Klara Volckaert}
\thanks{These two authors contributed equally}
\affiliation{Department of Physics and Astronomy, Interdisciplinary Nanoscience Center (iNANO), Aarhus University, 8000 Aarhus C, Denmark}

\author{Dmytro Kutnyakhov}
\affiliation{Deutsches Elektronen-Synchrotron DESY, Notkestr. 85, 22607 Hamburg, Germany}

\author{Steinn Ymir Agustsson}
\affiliation{Johannes Gutenberg-Universit\"at, Institut f\"ur Physik, 55099, Mainz, Germany}

\author{Kevin B\"uhlmann}
\affiliation{Department of Physics, Laboratory for Solid State Physics, ETH Z\"urich, Otto-Stern-Weg 1, 8093 Z\"urich, Switzerland}

\author{Federico Pressacco}
\affiliation{Deutsches Elektronen-Synchrotron DESY, Notkestr. 85, 22607 Hamburg, Germany}

\author{Michael Heber}
\affiliation{Deutsches Elektronen-Synchrotron DESY, Notkestr. 85, 22607 Hamburg, Germany}

\author{Siarhei Dziarzhytski}
\affiliation{Deutsches Elektronen-Synchrotron DESY, Notkestr. 85, 22607 Hamburg, Germany}

\author{Yves Acremann}
\affiliation{Department of Physics, Laboratory for Solid State Physics, ETH Z\"urich, Otto-Stern-Weg 1, 8093 Z\"urich, Switzerland}

\author{Jure Demsar}
\affiliation{Johannes Gutenberg-Universit\"at, Institut f\"ur Physik, 55099, Mainz, Germany}

\author{Wilfried Wurth}
\thanks{Deceased.}
\affiliation{Deutsches Elektronen-Synchrotron DESY, Notkestr. 85, 22607 Hamburg, Germany}
\affiliation{Center for Free-Electron Laser Science CFEL, Hamburg University, Luruper Chausee 149, 22761 Hamburg, Germany}

\author{Charlotte E. Sanders}
\affiliation{STFC Central Laser Facility, Research Complex at Harwell, Harwell Campus, Didcot OX11 0QX, United Kingdom}

\author{Philip Hofmann}
\email{philip@phys.au.dk}
\affiliation{Department of Physics and Astronomy, Interdisciplinary Nanoscience Center (iNANO), Aarhus University, 8000 Aarhus C, Denmark}

\date{\today}
\begin{abstract}
X-ray photoelectron diffraction is a powerful tool for determining the structure of clean and adsorbate-covered surfaces. Extending the technique into the ultrafast time domain will open the door to studies as diverse as the direct determination of the electron-phonon coupling strength in solids and the mapping of atomic motion in surface chemical reactions. Here we demonstrate time-resolved photoelectron diffraction using ultrashort soft X-ray pulses from the free electron laser FLASH. We collect Se~3d photoelectron diffraction patterns over a wide angular range from optically excited \ce{Bi2Se3} with a time resolution of 140~fs. Combining these with multiple scattering simulations allows us to track the motion of near-surface atoms within the first 3~ps after triggering a coherent vibration of the A$_{1g}$ optical phonons. Using a fluence of 4.2~mJ/cm$^2$ from a 1.55~eV pump laser, we find the resulting coherent vibrational amplitude in the first two interlayer spacings to be on the order of 1~pm.
\end{abstract}
\maketitle

The study of ultrafast dynamics in solids has not only provided an unprecedented insight into interactions between different degrees of freedom~\cite{Kirilyuk:2010vd,Giannetti:2016uf,Buzzi:2018wu}, it has also introduced methods for preparing entirely new transient quantum states~\cite{Torre:2021wg}.
Typical experiments use optical and UV lasers for pump-probe experiments, revealing the time-resolved electronic and optical properties for a wide range of solids, from bulk materials to two-dimensional layers~\cite{Maiuri:2020ve,Bovensiepen:2012tm}.
Ultrafast structural determination, on the other hand, is far less developed.
Time-resolved (TR) variants of traditional X-ray diffraction (XRD) harbor a huge potential~\cite{Lindenberg:2017aa,Buzzi:2018wu} but their use is restricted by several factors.
One is the scarcity of ultrafast X-ray sources, something that is beginning to change.
Another is the bulk-sensitivity of XRD that is ill-matched with the more surface-localized optical pump excitation or, indeed, the extreme surface sensitivity of time- and angle-resolved photoelectron spectroscopy (TR-ARPES) used to study the electron dynamics that accompanies any structural changes.
For instance, the phonon-driven modulations in the electronic structure measured by TR-ARPES can give a direct and detailed insight into the electron-phonon coupling of the system~\cite{Gerber:2017aa,Giovannini:2020aa} but only under the condition that the phonon-induced structural changes \emph{at the surface} are precisely known.
 
\begin{figure}
  \includegraphics[width=0.5\textwidth]{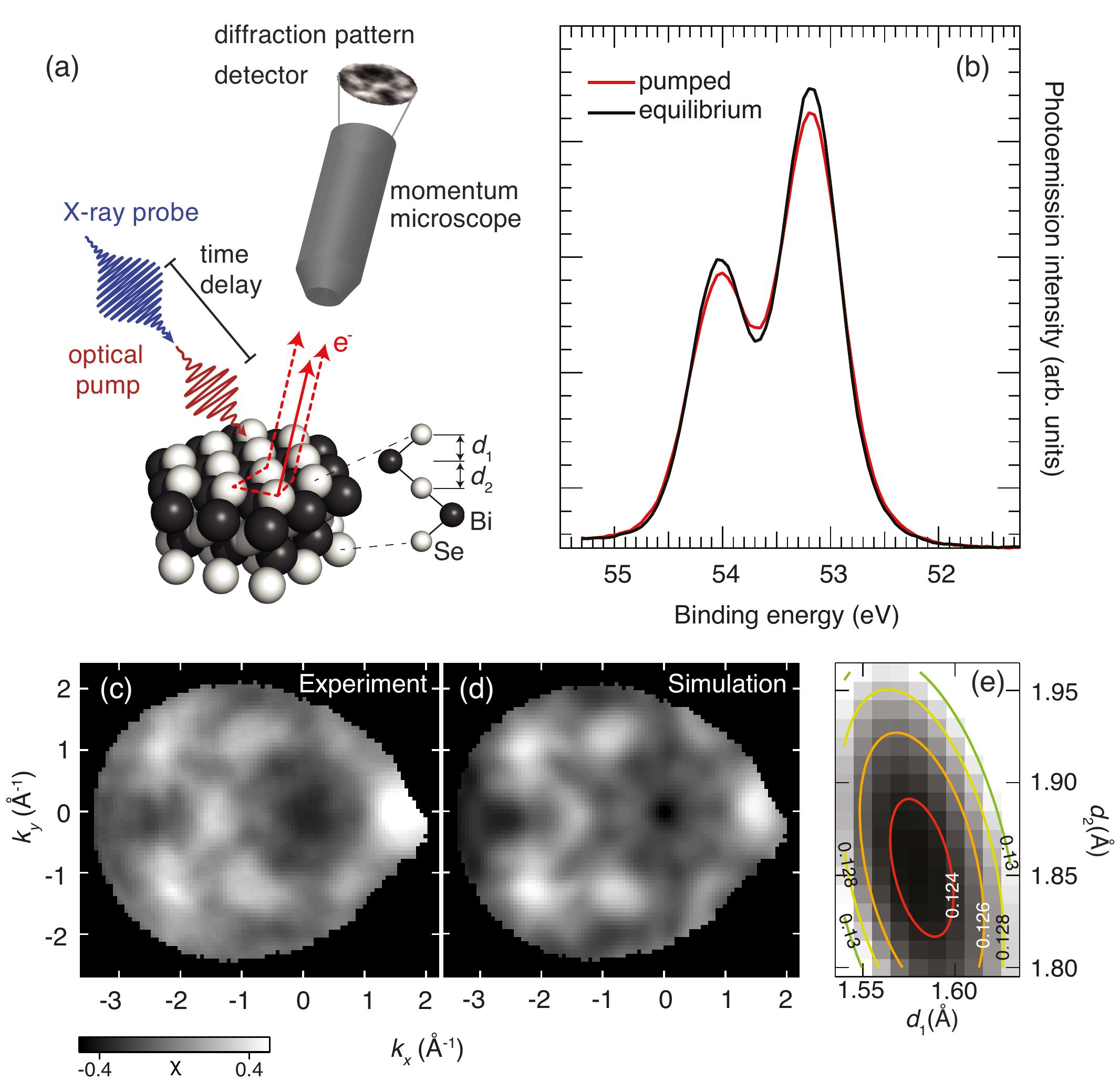}\\
  \caption{(Color online) (a) Principle of TR-XPD.
    A solid is excited with a short low-energy light pulse, followed by a soft X-ray pulse that gives rise to photoemission from core levels.
    The core level photoemission intensity shows an angular distribution that is given by the coherent interference of the photoelectrons' partial waves reaching the detector along different path ways, either directly (solid line) or via multiple-scattering events (dashed lines).
    (b) Se~3d core level spectrum before and after optical excitation.
    (c) Se~3d modulation function, obtained for integrated times of $t<$-0.075~ps.
    The scale bar applies to both panels c and d.
    (d) Calculated modulation function for the best-fit structure.
    (e) Reliability factor for the comparison between experimental and theoretical modulation functions as a function of the first two interlayer distances $d_1$ and $d_2$.
    The contours represent a fit to a two-dimensional polynomial.
  }
  \label{fig:1}
\end{figure}

A promising complementary technique to TR-XRD is TR-X-ray photoelectron diffraction (TR-XPD).
Static XPD is based on X-ray photoelectron spectroscopy (XPS) and inherits the chemical resolution and surface sensitivity from this technique:
It should thus be possible to measure atomic displacements at the very surface of a solid, where they are most relevant for a comparison to TR-ARPES data.
The principle of TR-XPD is illustrated in Fig.~\ref{fig:1}(a).
A solid is excited by an optical pump pulse, followed by an X-ray pulse leading to the emission of a core electron.
This electron can reach the detector directly or along different scattering pathways, and the measured intensity results from the coherent superposition of the direct and scattered wave field amplitudes.
The diffraction pattern measured at the detector is a fingerprint of the local atomic arrangement around the emitting atom and the structure can be determined by a comparison to multiple-scattering calculations~\cite{Woodruff:1992aa,Woodruff:1994ab,Osterwalder:1995vf,Hofmann:1994ae}.
Unlike XRD, XPD does thus not rely on long-range order and it has been used to study both long range and local phenomena such as surface relaxations~\cite{Kuznetsov:2015ab} and geometric changes during surface reactions~\cite{BAO:1994aa}.
TR-XPD using a pulsed X-ray source thus holds promise for real-time investigations for a wide range phenomena.
In fact, first TR-XPD demonstrations have been given for simple systems such as aligned molecules~\cite{Boll:2014up,Kastirke:2020aa}.
The effect has also been observed for solid surfaces~\cite{Greif:2016aa,Ang:2020aa}, but a rigorous structural determination based on such data is still missing.

Here we demonstrate the power of TR-XPD to track the ultrafast dynamic changes in the surface structure upon the excitation of coherent phonons on the surface of the topological insulator \ce{Bi2Se3}.
Symmetric A$_{1g}$ coherent optical phonons can be launched~\cite{Qi:2010ur,Kumar:2011ue} and controlled~\cite{Hu:2018ab} by ultrashort optical laser pulses.
The A$^{1}_{1g}$ phonon relevant to the present study has a frequency on the order of 2~THz and decays with a time constant of $\approx$3~ps~\cite{Hu:2018ab}.
The excitation of coherent phonons leads to small time-dependent variations in the binding energy of the electronic surface and bulk states, as measured by TR-ARPES~\cite{Sobota:2014aa}.
The bulk states show a binding energy oscillation consistent with the A$^{1}_{1g}$ mode, whereas the surface state energy modulation can only be described by superimposing a second vibration of a slightly lower frequency (2.05~THz instead of 2.23~THz)~\cite{Sobota:2014aa}.
This is consistent with a softening of force constants near the surface and the presence of a surface-localized vibrational mode~\cite{Ruckhofer:2020vg}.

TR-XPD experiments on the Se~3d core levels of \ce{Bi2Se3} have been performed with the HEXTOF experimental station at the PG2 beamline of the free electron laser FLASH~\cite{Kutnyakhov:2020aa,Gerasimova:2011aa}.
Samples were cleaved in vacuum and held at room temperature during the measurements.
The pump photon energy and fluence were 1.55~eV and $\approx$~4.2~mJcm$^{-2}$, respectively.
The probe photon energy was 113~eV.
Major challenges for TR-XPD are the need to collect high-quality TR-XPS spectra, not only integrated over the detector but for all emission angles individually.
This necessitated a total data collection time of $\approx$~19~hours.
After correcting for jitter from FLASH, the time resolution was 140~fs.
For these and other experimental details, see Appendix.

\begin{figure}
  \includegraphics[width=0.488\textwidth]{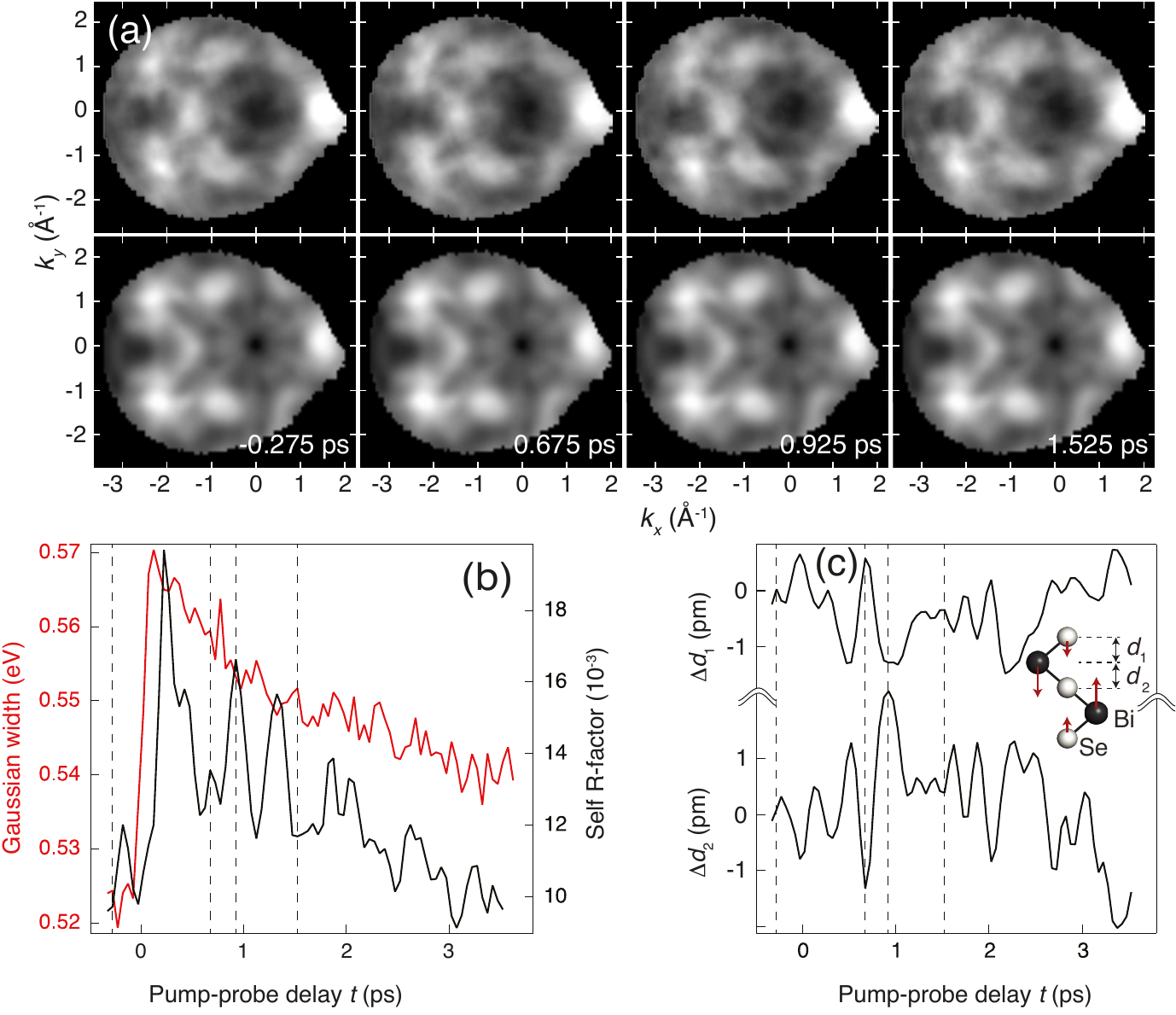}\\
  \caption{(Color online) Time-dependent structural analysis.
    (a) Top row: TR-XPD patterns integrated over an interval of 60~fs around selected time delays.
    Lower row: Best-fit calculated patterns, fitting only the parameters $\Delta d_1(t)$ and $\Delta d_2(t)$, i.e. the deviation of the interlayer spacings from the equilibrium position.
    (b) Black: self-$R$-factor of TR-XPD patterns with equilibrium pattern in Fig~\ref{fig:1}(c), red: Gaussian width of Se~3d$_{5/2}$ line.
    (c) Time-dependent interlayer distance changes $\Delta d_1(t)$ and $\Delta d_2(t)$.
    The inset shows the displacement for an A$^1_{1g}$ phonon (red arrows).
    The dashed lines in (b) and (c) indicate the time delays for the images in (a).
  }
  \label{fig:2}
\end{figure}

Figure~\ref{fig:1}(b) shows the effect of optical pumping on the Se~3d core level line shape.
There is a small but clear difference between the the spectrum collected before the arrival of the pump pulse (black) and the spectrum at peak excitation (red).
The spectra could be fitted by a Doniach-\v Sunji\'c line shape~\cite{Doniach:1970aa} with a very small asymmetry (see Appendix).
Pumping the material leads to a small increase in the Gaussian line width that could be used to estimate the experimental time resolution (see Figure~\ref{fig:2}(b)).
However, the effect is much smaller than in other materials~\cite{Dendzik:2020aa,Curcio:2021uh}.

TR-XPD patterns were obtained from the fits as the area under the \ce{Se}~3d peak, using Gaussian binning of the photoemission intensity as described in the Appendix.
Instead of representing the XPD pattern as the $\mathbf{k}$-dependent \ce{Se}~3d photoemission intensity $I(\mathbf{k})$, we use the so-called modulation function defined by $\chi(\mathbf{k})=({I(\mathbf{k})-I_0(\mathbf{k})})/{I_0(\mathbf{k})}$, where $I_0(\mathbf{k})$ is a two-dimensional polynomial~\cite{Woodruff:1992aa}.
This modulation function is displayed in Fig.~\ref{fig:1}(c) for the photoemission intensity in the unexcited state:
It is obtained by integrating the data at time-delays $t<$-0.075~ps, where $t=0$ corresponds to the highest temporal overlap of the pump and probe pulses; and negative values correspond to the probe pulse arriving before the pump pulse.

Before addressing the time-resolved atomic motion after an optical excitation, we determine the quasi-static geometric structure of \ce{Bi2Se3} by comparing the equilibrium XPD pattern in Fig.~\ref{fig:1}(c) to multiple scattering calculations performed by the software package for electron diffraction in atomic clusters~\cite{Garcia-de-Abajo:2001aa}.
The calculated modulation function for the optimized structural and non-structural parameters is given in Fig.~\ref{fig:1}(d)~(see Appendix).
The agreement with the experimental result is excellent.
To quantify this, the two modulations functions can be compared by the reliability factor 
\begin{equation}
R=\frac{\sum_i (\chi_{e,i} - \chi_{t,i})^2}{ \sum_i (\chi_{e,i}^2 + \chi_{t,i}^2 )},
\label{eq:rf}
\end{equation}
where $\chi_{e,i}$ and $\chi_{t,i}$ are the experimental and theoretical modulation functions for the $i$'th $\mathbf{k}$-point, respectively.
We find a very low value of $R=$0.12.
In order to reach this agreement, the structural parameters were initially fixed to the bulk values for \ce{Bi2Se3}~\cite{Nakajima:1963aa}.
An optimization was limited to the first two interlayer distances $d_1$ and $d_2$, as defined in Figure~\ref{fig:1}(a).
$R$ is plotted as a function of $d_1$ and $d_2$ in Figure~\ref{fig:1}(e).
The values of $R$ are fitted to a two-dimensional polynomial in order to determine the optimum parameters $d_1$ and $d_2$ with a higher precision than given by the grid of $R$ calculations.
The best agreement between experiment and calculation is found for $d_1=$1.579$\pm$0.050~{\AA} and $d_2=$1.866$\pm$0.100~{\AA}.
The uncertainties are determined from the variance of the $R$-factor~\cite{Bana:2018aa}.
With respect to the bulk structural parameters, there is a small inward relaxation of the first interlayer spacing, in excellent agreement with a previous surface structure determination by surface XRD and low energy electron diffraction~\cite{Reis:2013aa}, and in fair agreement with a static XPD investigation~\cite{Kuznetsov:2015ab}.
The second interlayer spacing also shows a small inward relaxation whereas in previous experiments~\cite{Reis:2013aa,Kuznetsov:2015ab} it was found to be very similar to the bulk value.
For more details on the multiple scattering simulations, see Appendix.

The average structure is now taken as a starting point to determine the time-dependent changes of $d_1$ and $d_2$ after the optical excitation.
XPD patterns at selected points in time are shown in the top row of Figure~\ref{fig:2}(a).
Figure~\ref{fig:2}(b) shows a basic characterization of the time-dependent data with the vertical dashed lines corresponding to the time points chosen for the XPD patterns in panel (a).
The black line is a self-$R$ factor between experimental data alone, comparing the measured modulation function at each time delay with the static modulation function in Figure~\ref{fig:1}(c).
Note that this self-$R$ factor is defined as in equation (\ref{eq:rf}), merely exchanging the modulation functions which are to be compared.
This allows us to judge -- from experimental data alone -- how much the pattern changes after the optical excitation.
The self-$R$ factor shows a rapid increase shortly after $t=0$, followed by a gradual decay as the surface relaxes back to its equilibrium structure (there can be additional changes of the overall lattice constant due to the induced carriers over a much larger time scale than investigated here~\cite{Kim:2021vf}).
The decay is superimposed with a pronounced oscillatory structure, indicating that the path back to equilibrium is accompanied by periodic structural changes, such as optical phonons.
Figure~\ref{fig:2}(b) also gives the time-dependent Gaussian width of the Se~3d$_{5/2}$ peak.
This shows step-wise broadening at $t=0$, followed by a gradual narrowing.
The width of the step establishes the time resolution of the experiment.
It is clearly significantly shorter than the period of the observed oscillations.

A quantitative structural analysis is now performed by comparing the time-resolved diffraction patterns to multiple scattering simulations like the one in Figure~\ref{fig:1}(d), while optimizing the structural parameters ($d_1$ and $d_2$) for each time delay, using a two-dimensional interpolation of $R$ as in Figure~\ref{fig:1}(e).
The resulting changes for the two interlayer distances with respect to the equilibrium values ($\Delta d_1$ and $\Delta d_2$) are shown in Figure~\ref{fig:2}(c) and the best fits to the measured XPD patterns are given in the lower row of Figure~\ref{fig:2}(a).
Both $\Delta d_1$ and $\Delta d_2$ show clear oscillations with an amplitude on the order of 1~pm, especially immediately after excitation and towards the end of the explored time interval.
The period of the oscillation is about 500~fs, consistent with the expected excitation of an A$^{1}_{1g}$ optical phonon.
The structural changes are also in line with what is expected for an A$^{1}_{1g}$ phonon.
As seen in the inset, this phonon mode involves the simultaneous movement of the two outer layers (Bi and Se) in a quintuple layer (QL) with respect to the static central Se layer.
The changes in $d_1$ and $d_2$ should thus be correlated and be either in-phase or out-of-phase, depending on the relative size of the movements in the first and second layer atoms.
The in-phase (out-of-phase) motion corresponds to a larger (smaller) displacement of the outer Se atom from the central atom than the displacement of the Bi atom.
Figure~\ref{fig:2}(c) shows the displacements to be approximately out-of-phase.

The structural changes should be treated with some caution.
As seen in Fig.~\ref{fig:2}(a), the changes in the diffraction pattern are very small.
The changes in $d_1$ and $d_2$ have the expected order of magnitude~\cite{Johnson:2013wc,Gerber:2017aa} but they are much smaller than the uncertainties for the determination of the static values of the interlayer distances.
We can thus expect that some of the movements seen in Figure~\ref{fig:2}(c) are due to noise.
On the other hand, we need to keep in mind that uncertainties stated for the static structural parameters also account for systematic errors, for example from oversimplifications in the multiple scattering simulations.
Such effects limit the overall reachable agreement (and hence, the uncertainties) but they are identical for the calculations performed at each point in time, suggesting that the changes in $d_1$ and $d_2$ should be more reliable than their absolute values.
The purely statistical uncertainties due to noisy data can be estimated from the structural fluctuations for negative time delays in Figure~\ref{fig:2}(c) and these are much smaller than the static uncertainties and also smaller than the pronounced structural oscillations at later time delays.
Finally, a structural change is supported by the oscillatory shape of the time-dependent $R$-factor in Figure~\ref{fig:2}(b) and the fact that both the oscillation frequency and the movements in $d_1$ and $d_2$ are consistent with an excited A$^{1}_{1g}$ optical phonon.

\begin{figure}
  \includegraphics[width=0.5\textwidth]{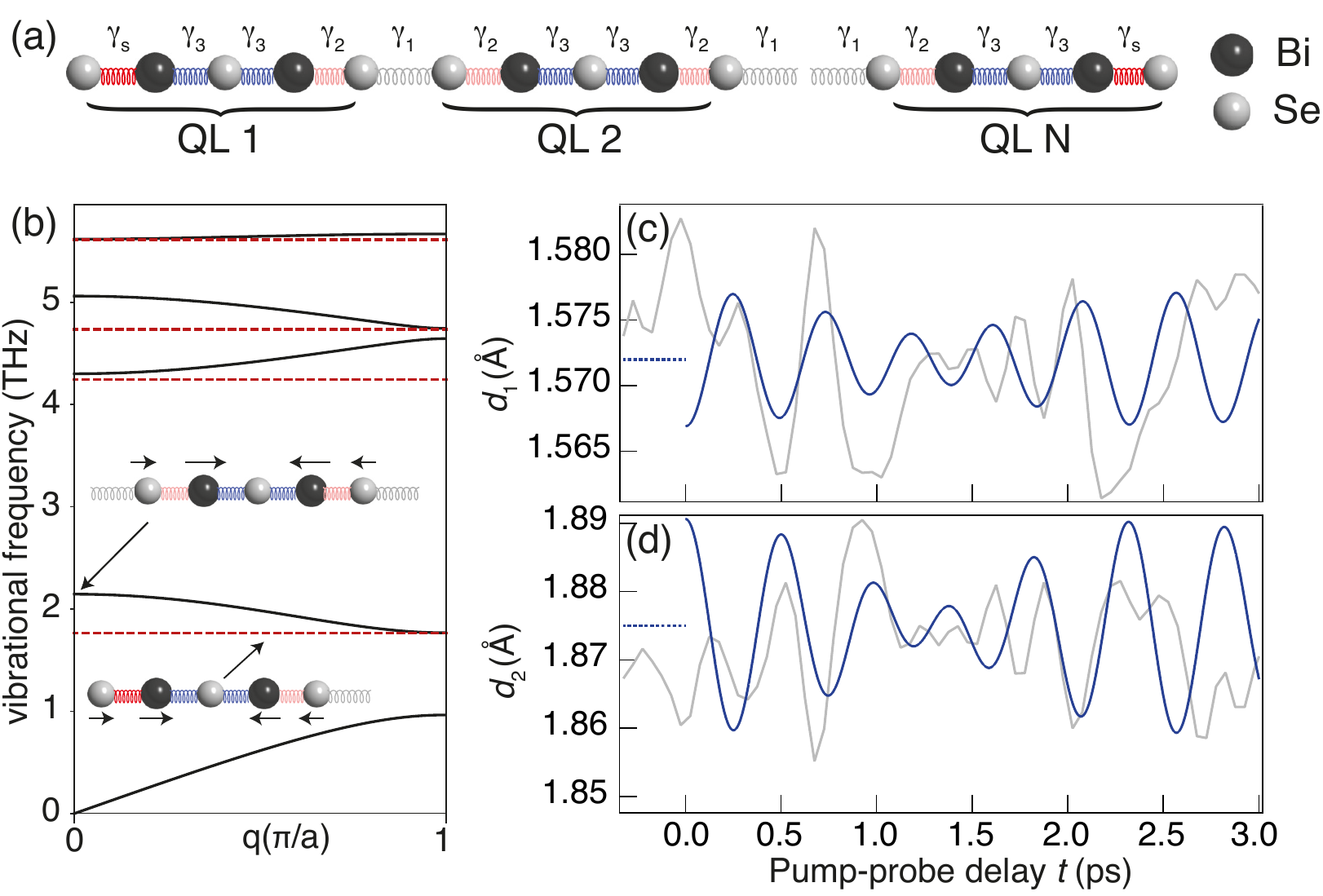}\\
  \caption{(Color online) (a) One-dimensional linear chain model with five atoms per unit cell and nearest neighbor interactions.
    The bulk is described by three force constants $\gamma_{1,2,3}$.
    The first layer force constant is permitted to be different ($\gamma_{1s}$).
    (b) Phonon dispersion curves for the bulk (black solid lines) and the surface-localized modes (red dashed lines).
    (c) and (d) Dark blue line: Time-dependent displacement of the first and second atomic spacing in the one-dimensional chain, respectively, when exciting both the bulk A$^{1}_{1g}$ mode and the A$^{1}_{1g}$ derived end mode for $d_1$ (c) and $d_2$ (d).
    The dashed lines before $t=0$ represent the equilibrium values of the linear chain model.
    Light grey line: experimental displacements from Figure~\ref{fig:2}(c).
  }
  \label{fig:3}
\end{figure}

For a more detailed understanding of the observed structural changes, we introduce a minimal one-dimensional model to study time-dependent interlayer distances near the surface.
The model is illustrated in Figure~\ref{fig:3}(a).
We consider a linear chain of atoms with a Se-Bi-Se-Bi-Se QL as basis.
The model has three different spring constants, $\gamma_2$ and $\gamma_3$ within a unit cell and a weaker $\gamma_1$ between the unit cells, representing the van der Waals forces between the QLs of \ce{Bi2Se3}.
We adjust the force constants such that the phonon dispersion is similar to that of \ce{Bi2Se3} in the $\Gamma$-Z direction of the bulk Brillouin zone (see Figure~\ref{fig:3}(b))~\cite{Zhu:2011ab,Ruckhofer:2020vg}.
The A$^{1}_{1g}$ phonon mode corresponds to the lowest optical branch at the zone centre.
Its displacement pattern is shown in the inset.
In the model, the vibrational amplitude for the second layer Bi atoms is higher than that of the first layer Se atoms, consistent with an out-of-phase motion of $d_1$ and $d_2$.
Using the bulk force constants does not give rise to distinct vibrations at the end of the chain.
We therefore introduce a localized end mode by choosing a softer force constant $\gamma_s$ at the end of the chain (reduced by 10\% with respect to $\gamma_2$).
This results in several end-localized modes with frequencies that do not appear in the bulk continuum (dashed red lines in Figure~\ref{fig:3}(b)).
The displacement pattern of the A$^{1}_{1g}$-derived end mode is also given as an inset.
Its slight asymmetry is caused by the the softer $\gamma_s$ and the missing spring at the end of the chain.

Using a superposition of the A$^{1}_{1g}$ bulk and end modes, it is possible to achieve a qualitative agreement with the experimentally observed displacements, as shown in Figure~\ref{fig:3}(c) and (d).
In this picture, the reduced vibrational amplitude in the middle of the investigated delay time window is the result of a beating pattern of the two modes.
The very simple model can of course not be expected to give a quantitative description of $d_1(t)$ and $d_2(t)$ and there is some freedom to choose the best parameters for reproducing the experimental displacements (the relative excitation strength and the phase between the modes, see Appendix).

In summary, we have demonstrated the use of TR-XPD for an ultrafast tracking of the surface atomic structure after the excitation of a coherent phonon in \ce{Bi2Se3}.
The coherent vibrational amplitudes for the pump fluence used here are on the order of 1~pm, allowing the calculation of the deformation potential when combined with TR-ARPES experiments.
An important challenge for TR-XRD is the need for high-quality XPS spectra when binning the collected data in both $\mathbf{k}$ and time.
This leads to very long data acquisition times and, in our case, limits the length of the delay time interval that can be studied and the precision of the obtained structural parameters.
The bottleneck in the data acquisition is vacuum space charge~\cite{Schonhense:2018aa} and a desirable characteristics of future free electron laser sources will be a much increased pulse repetition rate.
This will allow similar studies not only to take data sets of better quality and over longer time delays but also at multiple photon energies, drastically increasing the precision of the structural determination.
Eventually, the chemical and local sensitivity of TR-XPD could be exploited, e.g., to probe the detailed atomic motion in specific coherently excited vibrational modes in molecules, representing a time-resolved version of the previously suggested use of XPD to measure the time-averaged probability distribution of atoms~\cite{Hofmann:1996aa}.

\begin{acknowledgments}
This work was supported by VILLUM FONDEN via the Centre of Excellence for Dirac Materials (Grant No. 11744) and the Independent Research Fund Denmark (Grant No. 1026-00089B).
This research was carried out at FLASH at DESY, a member of the Helmholtz Association.
The research leading to these results has been supported by the project CALIPSOplus under the Grant Agreement 730872 from the EU Framework Programme for Research and Innovation HORIZON 2020, by the Deutsche Forschungsgemeinschaft (DFG) within the framework of the Collaborative Research Centre SFB 925 - 170620586 (project B2) and within TRR 173-268565370 (project A05).
We thank Holger Meyer and Sven Gieschen and Harald Redlin for experimental support, as well as Gerd Sch\"onhense, Davide Campi, Anton Tam\"ogl and Wolfgang Ernst for helpful discussions.
\end{acknowledgments}

\section{Appendix}

\subsection{Experimental Details}
\subsubsection{Pump and free electron laser pulse characteristics}

The free electron laser (FEL) light pulses provided by the FLASH FEL at the PG2 beamline were provided with a structure of 425 pulses within each pulse train, of which the last 25 were unpumped for reference data.
The pulse trains arrived with a frequency of 10~Hz, while the intra-pulse frequency was 1~MHz.
The pulses were produced with a fundamental wavelength of 9.3~nm, and they were monochromatized using an SX-700 type monochromator~\cite{Martins:2006aa}.
The selected photon energy was 113~eV, and the polarization on the sample was linear p-type.

\begin{figure}[htb]
  \includegraphics[width=0.5\textwidth]{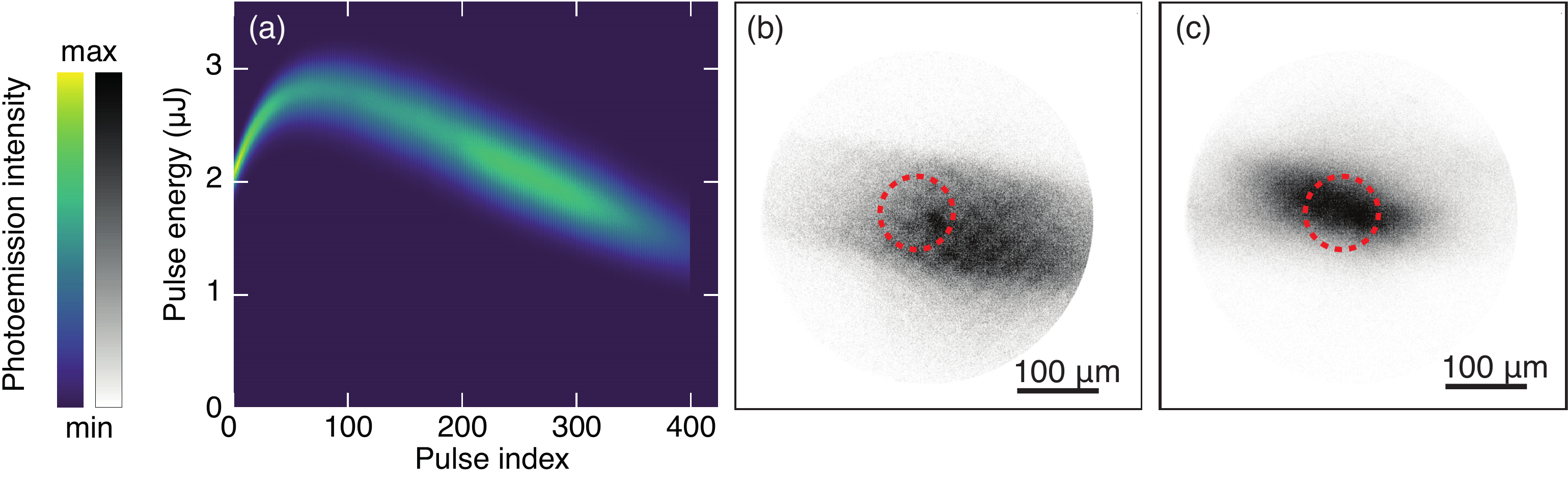}\\
  \caption{
    (a) Photoemission intensity as a function of pump pulse energy measured by the optical diode and pulse index.
    The color scale shows the number of photoelectrons detected for each bin.
    (b) The FEL beam shape on the sample surface as seen by its photoemission intensity in real space.
    (c) The multiphoton photoemission intensity as excited by the optical pump beam.
    The dashed red circles in (b) and (c) mark the region selected by the aperture.
  }
  \label{fig:S1}
\end{figure}

The intensity of the FEL pulses was attenuated until only about 2.1 photoelectrons were detected per pulse, on average.
This was done to minimize both multi-hit artifacts on the momentum microscope dual 4-quadrant delay line detector, and to minimize FEL-induced space-charge effects~\cite{Schonhense:2015ab,Schonhense:2018aa,Schonhense:2021ab}.
The beam shape of the FEL can be seen in Fig.~\ref{fig:S1}(b).

The pump pulses were provided by an optical parametric chirped pulse amplification laser~\cite{Redlin:2011aa}.
The energy was 1.55~eV and the polarization on the sample was linear s-type, to minimize laser-assisted photoemission spectral replicas.
The pump pulses were delivered to the sample collinear to the FEL beam, with an incidence angle of 68$^{\circ}$ off-normal.

The pump intensity was measured on a per-shot basis using a calibrated photodiode, and the intensity was recorded together with every detected photoelectron.
The average pulse energy was 1.2~\textmu J, but a substantial shot-to-shot variability of the pump intensity was present, as can be seen in Fig.~\ref{fig:S1}(a).

In order to calibrate the pump laser fluence, the laser profile was measured by using the momentum microscope in spatial microscopy mode, following the same approach as in Refs.~\cite{Curcio:2021uh} and~\cite{Dendzik:2020aa}.
This allowed to directly measure the pump laser area distribution on the sample surface (Fig.~\ref{fig:S1}(c)).
In estimating the pump fluence, we have also considered the effect of a 50~\textmu m aperture positioned in the first Gaussian image plane of the momentum microscope, and a multiphoton photoemission power law exponent of 4.
This analysis reveals an average pump fluence of 4.2~mJ/cm$^2$.

\subsubsection{FEL jitter correction}
In order to correct for the timing pulse-to-pulse jitter, several strategies have been employed.
The first correction is given by a bunch arrival monitor, which is present on the undulator beamline, and which gives the time delay of each FEL shot compared to a common clock.
This correction is capable of compensating for the shot-to-shot variability arising from the self amplified spontaneous emission nature of FLASH.
This correction improves the overall time resolution from 200~fs to 170~fs.

\begin{figure}[htb]
  \includegraphics[width=0.5\textwidth]{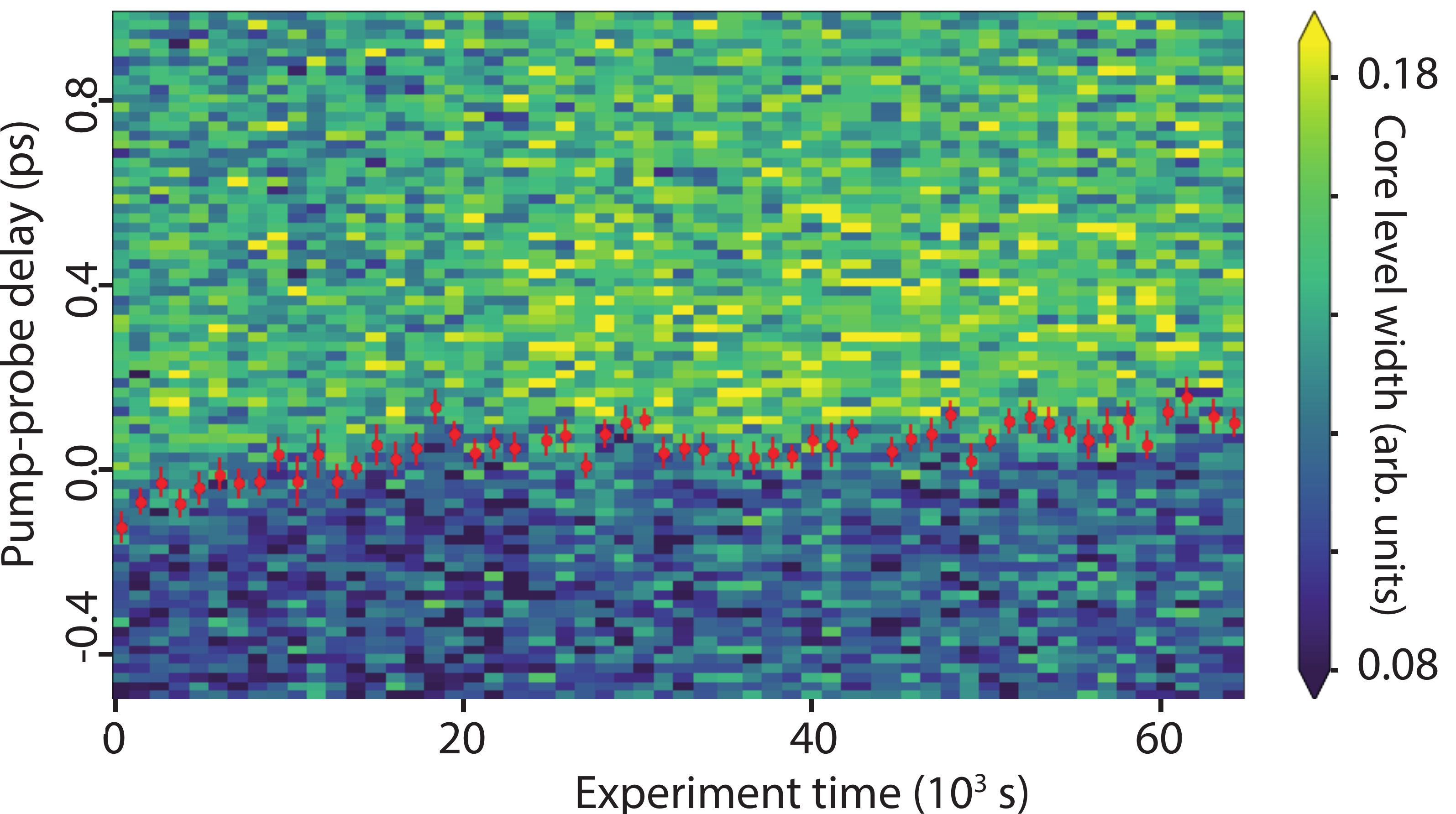}\\
  \caption{
    Simple measure of the core level peak width of \ce{Se}~3d and \ce{Bi}~5d as a function of pump-probe delay $t$ and running time of the experiment. 
    A drift of the characteristic peak broadening near $t=0$ can be observed and corrected for. 
    The red markers show the detected position $t=0$, with associated uncertainty.
  }
  \label{fig:S2}
\end{figure}

The second jitter correction is performed by measuring the broadening of the \ce{Se}~3d and \ce{Bi}~5d core levels as a function of pump probe delay and experiment time (see Fig.~\ref{fig:S2}). 
This broadening is quantified in a simple way by the spill-out of photoemission intensity from the center of the peaks to their tails. 
A fast increase in width of the core levels marks pump and probe temporal overlap and can effectively be used to track the ``time zero'' drifts over experiment time.
Based on this, we can compensate for slow drifts in the timing delays between pump and probe, and is performed with a time binning of 1200~s.
Since these slow drifts can be substantial, of the order of 100~fs, this correction gives a significant time resolution improvement, from 170~fs to 140~fs.

The ``time zero'' position was calibrated by rotating the pump laser polarization to p-type, so that laser assisted photoemission gives a clear photoemission signal 1.55~eV to lower binding energy of the \ce{Se}~3d core levels when pump and probe overlap in time.

\subsubsection{Binding energy axis calibration and correction}
The binding energy is not directly measured by the momentum microscope.
Rather, the time-of-flight of each photoelectron compared to a master clock trigger is recorded.
From this, the kinetic energy was calibrated following the approach of Refs.~\cite{Curcio:2021uh} and~\cite{Dendzik:2020aa}.
The absolute binding energy of the core levels was then calibrated using photoemission data collected at the SGM3 beamline of the synchrotron radiation source ASTRID2 at Aarhus University~\cite{Hoffmann:2004aa}.

\begin{figure}[htb]
  \includegraphics[width=0.5\textwidth]{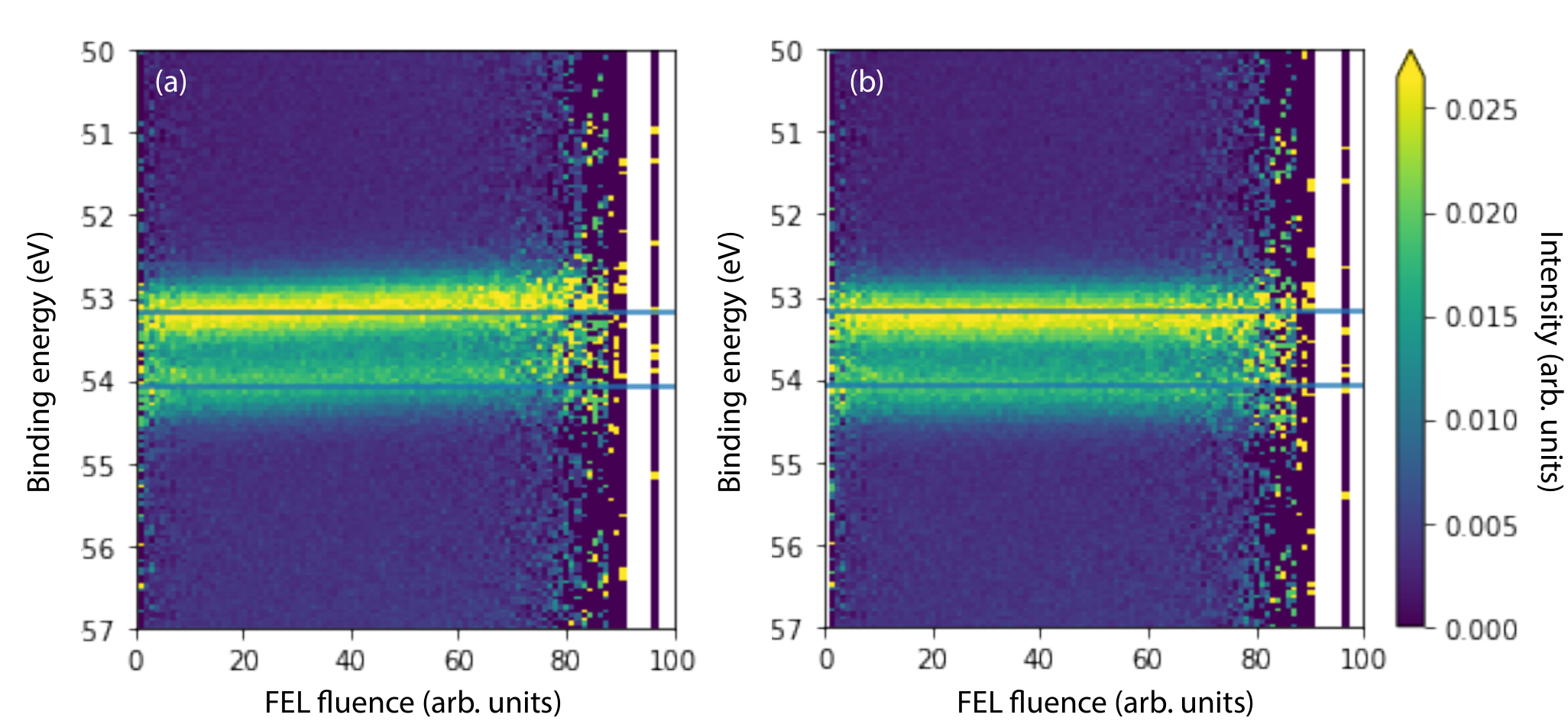}\\
  \caption{
    (a) Photoemission intensity in the \ce{Se}~3d core level region as a function of probe fluence, illustrating the effect of probe-induced space-charge.
    The horizontal blue lines are placed at the binding energies measured in photoemission studies performed at the SGM3 beamline of ASTRID2.
    (b) The corrected spectra.
    In both panels the data corresponds to the first 1000~s of acquisition time.
  }
  \label{fig:S3}
\end{figure}

Two additional corrections have been applied:
The first one considers a dependence of space-charge shifts on the FEL fluence.
As can be seen in Fig.~\ref{fig:S3} (a), the apparent binding energy of the core level peaks systematically depends on the FEL fluence.
This can be corrected individually for each detected photoelectron, leading to the result in Fig.~\ref{fig:S3} (b).
This has a small effect on energy resolution, but it eliminates possible artifacts due to correlations between probe pulse intensity and other parameters.

The final correction is the removal of a well-known kinetic energy distortion by the momentum microscope.
In momentum microscopes, there is a radial dependence of the photoelectron time-of-flight in parallel momentum space due to space-charge and geometric differences in travel distance of the photoelectrons~\cite{Schonhense:2018aa}.
A correction of this effect gives a dramatic improvement in energy resolution when integrating core levels over a (partial) detector area, such as in Fig.~1(b) of the main text.

\subsubsection{Parallel momentum axes calibration}
\begin{figure}[htb]
  \includegraphics[width=0.5\textwidth]{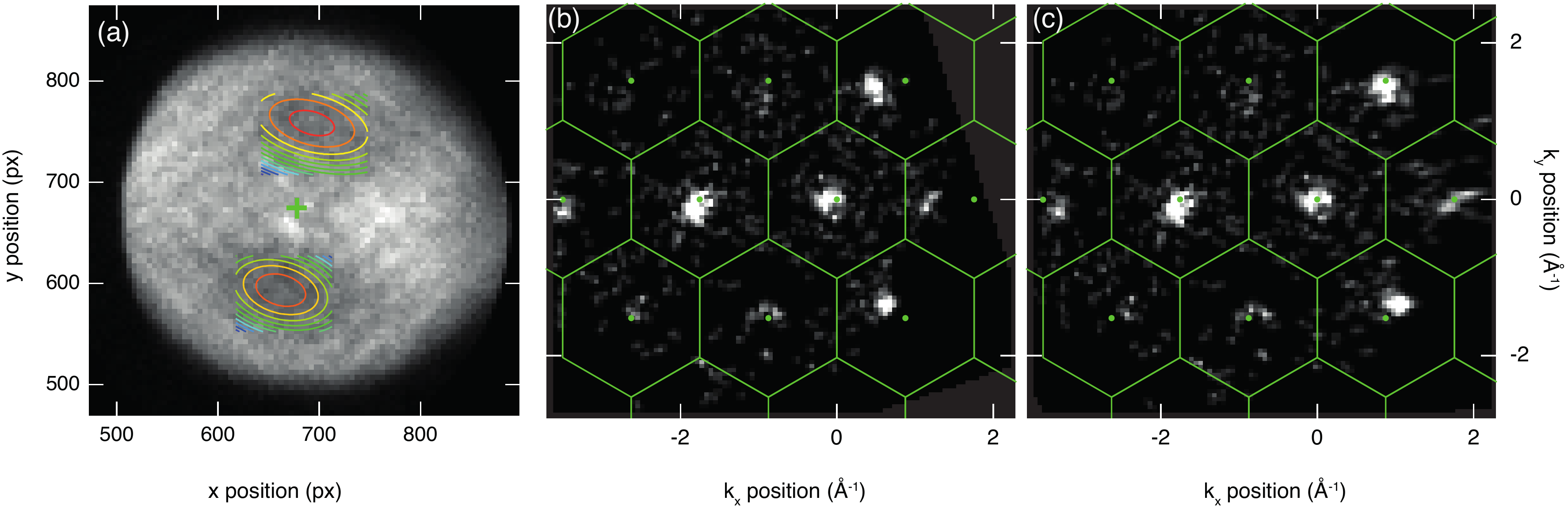}\\
  \caption{
    (a) Out-of-focus illumination pattern of the detector for an integrated slice (integrating from 51.95~eV to 51.45~eV, and from -250~fs to 3~ps), showing two 2-dimensional polynomial fits used to determine the center of distortion.
    The green cross marks the middle position between the minima of the fitting polynomials.
    (b) Photoemission at the Fermi level after conversion from pixel space to $k$-space but without the application of a distortion correction.
    The green lines and points represent the surface Brillouin zones and $\bar{\Gamma}$ points in the extended zone scheme.
    (c) Same data as in panel (b) but after applying the distortion correction.
  }
  \label{fig:S4}
\end{figure}

The detector of the momentum microscope gives the location of impinging electrons in pixel space and these locations need to be transformed to momentum space for a comparison with the multiple scattering calculations.
The transformation needs to include a correction of imaging distortions.
To this end, we are employing the so-called radial ``division model'', commonly used for correcting barrel or pincushion type distortions.
This is implemented by the following equations:
\begin{equation}
  \label{eq:eqS1}
  \begin{split}
  r_{k} &= \frac{r_{px}}{K_0+K_1r_{px}^2+K_2r_{px}^4}\\
  \theta_{k} &= \theta_{px} + \theta_{offset}
  \end{split}
\end{equation}
with
\begin{equation}
  \label{eq:eqS2}
  \begin{split}
    r_{px} &= \sqrt{(x_{px}-x_{0,px})^2+(y_{px}-y_{0,px})^2}\\
    \theta_{px} &= \tan^{-1}\left( \frac{y_{px}-y_{0,px}}{x_{px}-x_{0,px}}\right),\\
  \end{split}
\end{equation}
where $K_n$ are the distortion parameters; $r_k$ and $\theta_{k}$ are the radial coordinates in parallel momentum space; $r_{px}$ and $\theta_{px}$ are the polar coordinates in pixel space; $\theta_{offset}$ is an angular rotation parameter; $x_{px}$ and $y_{px}$ are the coordinates in pixel space, and $x_{0,px}$, $y_{0,px}$ is the origin of a ``center of distortion'' in pixel space.
This ``center of distortion'' is assigned to the the center of the out-of-focus photoemission intensity pattern on the low binding energy side of \ce{Se}~3d core level (see Fig.~\ref{fig:S4}(a)).
Its location is determined by fitting the intensity around the two visible dark areas with 2-dimensional polynomials of degree 2, and averaging the $x$ and $y$ coordinates of the minima.

The 3 $K_n$ parameters are determined maximizing overlap of the features in the diffraction pattern between experiment and multiple scattering simulations.
Note that the choice of $d_1$ and $d_2$ in the simulated diffraction patterns used for this purpose (within physically sensible bounds) has only a very small impact on the resulting $K_n$ parameters and an even smaller impact on the resulting optimized structural parameters.
This is expected from fact that the most dominant features in the diffraction pattern (and the most important features for the $R$-factor) are the pronounced forward scattering peaks, and that the position of these peaks is mostly determined by the structure of the very first Se layer.

This approach can also be tested on the valence band data.
To this end, Fig.~\ref{fig:S4}(b) shows the photoemission intensity at the Fermi level after a conversion to $k$-space but without applying the distortion correction.
Based on the electronic structure of Bi$_2$Se$_3$, one expects to observe a photoemission intensity maxima at the $\bar{\Gamma}$ points of the surface Brillouin zones in the extended zone scheme, either from the topological surface state or from the conduction band minimum.
Such maxima in the photoemission intensity are indeed visible.
Fig.~\ref{fig:S4}(c) shows the same data but with after the image distortion correction.
Clearly, the position of the intensity maxima is in much better agreement with the expected positions near the $\bar{\Gamma}$ points.
The $K_n$ values used for maximizing the overlap of the photoemission intensity at the Fermi level with the grid of $\bar{\Gamma}$ points are very similar to those used for the core level spectra, even if the sample alignment and lens settings are dramatically different.

After removing the distortion, the actual position of the features in momentum space needs to be calculated, requiring the determination of the normal emission direction, i.e. of the origin of momentum space.
Normal emission is found by exploiting the fact that, due to selection rules of photoemission, there is a low intensity spot around normal emission for the \ce{Se}~3d core levels (see Fig.~1(c) in the main text).
For a precise determination of the minimum's location, a 2-dimensional polynomial of degree 2 is fitted to the intensity pattern, yielding the minimum coordinates $x_{0,k}$ and $y_{0,k}$.
With this, the coordinates in momentum space are given by:
\begin{equation}
  \label{eq:eqS3}
  \begin{split}
    x_{k} &= x_{0,k}+r_{k}\cos(\theta_{k})\\
    y_{k} &= y_{0,k}+r_{k}\sin(\theta_{k})
  \end{split}
\end{equation}

\subsubsection{Data binning and data quality}
Photoelectron detection events are stored in a data table format, such that binning the data is required to give the 4-dimensional datasets.
As mentioned in the previous sections, the binning axes are a function of the acquisition axes, and consist in binding energy, pump-probe delay, and 2 lateral momentum axes.
In order to prevent artifacts due to rectangular bins, we employed an overbinning approach where binning in steps much smaller than our resolution and subsequent Gaussian convolution was used.

\begin{figure}[htb]
  \includegraphics[width=0.4\textwidth]{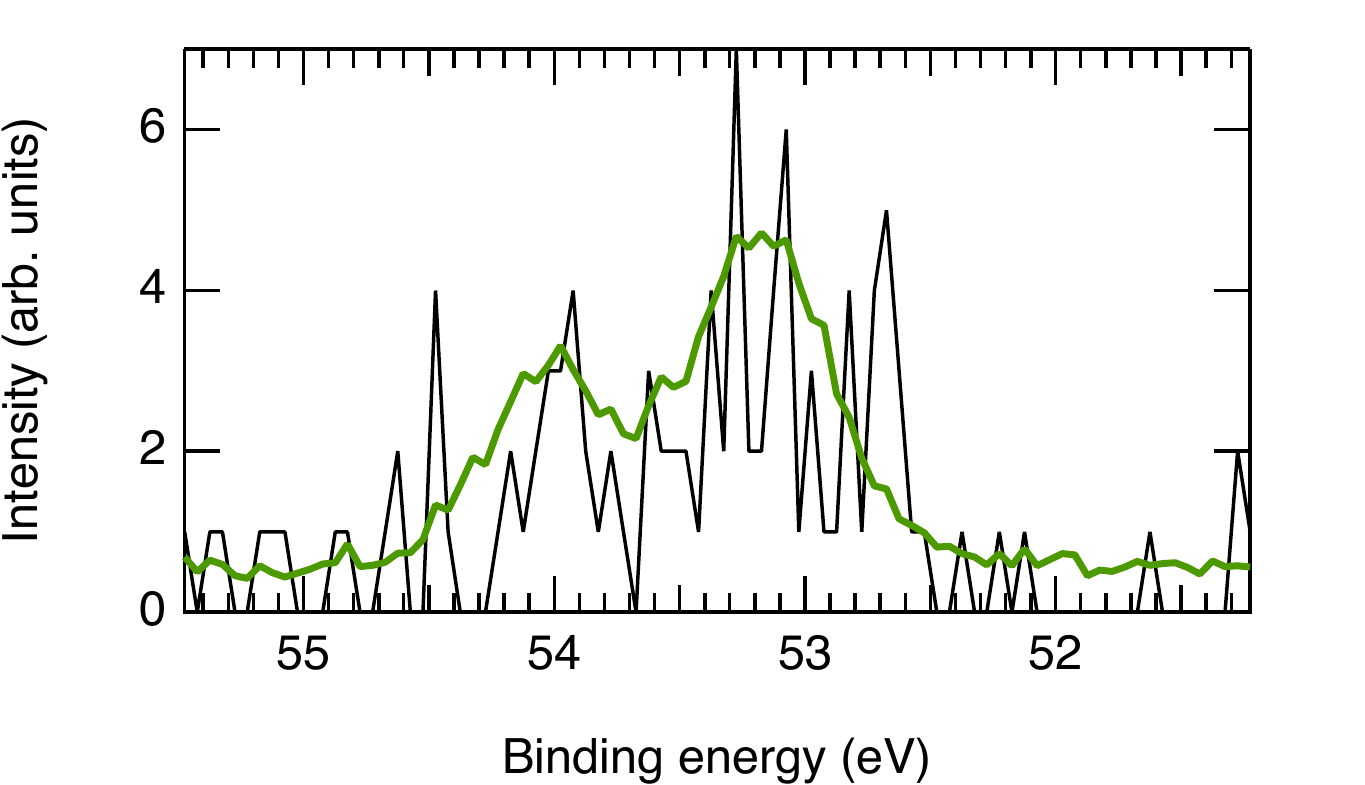}\\
  \caption{
    (a) Example \ce{Se}~3d overbinned spectrum (black) and Gaussian rebinned spectrum (green).
    Both spectra have been taken for a pump probe delay of -275~fs and a lateral momentum of -1.75~{\AA}$^{-1}$ along the $k_x$ direction, corresponding to a $\bar{\Gamma}$ point.
  }
  \label{fig:S5}
\end{figure}

We overbinned the pump probe delay and the momentum axes, with a resolution of 50~fs for the former and roughly 0.015~{\AA}$^{-1}$ for the latter (an example spectrum for a bin of this size is shown in Fig.~\ref{fig:S5}, black curve).
Then, a Gaussian kernel has been convoluted with the binned data, in order to maximize signal-to-noise ratio, while preserving experimental resolution.
The Gaussian widths for the 3 overbinned axes are 60~fs and 0.045~{\AA}$^{-1}$ for the pump probe delay and parallel momentum axes, respectively.
The binding energy axis was binned with a 50~meV energy step.

This procedure results in a 3-dimensional series of 1-dimensional core-level spectra (for a total of 4 dimensions).
For such spectra, the experimental resolution has not been affected by the binning procedure while maximizing the counts per bin.
An example of the resulting spectra is shown in Fig.~\ref{fig:S5} (green curve).

\subsubsection{XPS data fitting and XPD patterns}
Following the binning of the data, we fitted a sum of two Doniach-\v Sunji\'c functions and a linear background to the 4-dimensional pump-probe delay and lateral momentum dependent \ce{Se}~3d spectral series.
The fitting range was from 54.85~eV to 51.25~eV, and the equilibrium optimized fitting parameters (taking only negative time delays) are summarized in Table~\ref{tab:S1}.

\begin{table}[htb]
  \begin{center}
    \begin{tabular}{@{\hspace{1em}}l@{} @{\hspace{1em}}c@{} @{\hspace{1em}}c@{}} 
      \hline
                          & \ce{Se}~3d$_{5/2}$      & \ce{Se}~3d$_{3/2}$ \\
      \hline
      Lorentzian width    & 0.25$\pm$0.01~eV    & 0.23$\pm$0.03~eV \\ 
      Asymmetry parameter & 0.03$\pm$0.015     & 0.03$\pm$0.02 \\ 
      Gaussian width      & 0.48$\pm$0.01~eV    & 0.48$\pm$0.01~eV \\ 
      Intensity           & 7.7$\pm$0.4        & 4.7$\pm$0.4 \\ 
      Binding energy      & 53.168$\pm$0.005~eV & 54.03$\pm$0.01~eV \\
      \hline
    \end{tabular}
  \end{center}
  \caption{
    Fitting parameters used for the two Doniach-\v Sunji\'c peaks fitted to the \ce{Se}~3d spectra.
  }
  \label{tab:S1}
\end{table}

In all the fits, the Gaussian widths and asymmetry parameters of the two spin-orbit split components were constrained to be equal.
This line shape model was then fit to the pump-probe delay and lateral momentum dependent series of spectra.
Only the common Gaussian widths, the binding energies (with fixed spin-orbit splitting), and the intensities were allowed to vary.
The observed changes in binding energy were very small.

Once every spectrum in the time- and momentum- series has been analyzed, it is possible to extract the photoelectron diffraction patterns for each time delay.
These are generated by measuring the fitting function peak area for each spectrum.
The patterns have been normalized by the background intensity found under the peaks, because the detector and the FEL give a non uniform ``illumination'' that is hard to measure by other means, since it depends on the specific sample alignment and lens settings.
This is justified since the closest core level on the low binding energy side is \ce{Bi}~5d$_{5/2}$, which is about 27~eV away, and such intensity has retained negligible photoelectron diffraction angular information.
\begin{figure}[htb]
  \includegraphics[width=0.5\textwidth]{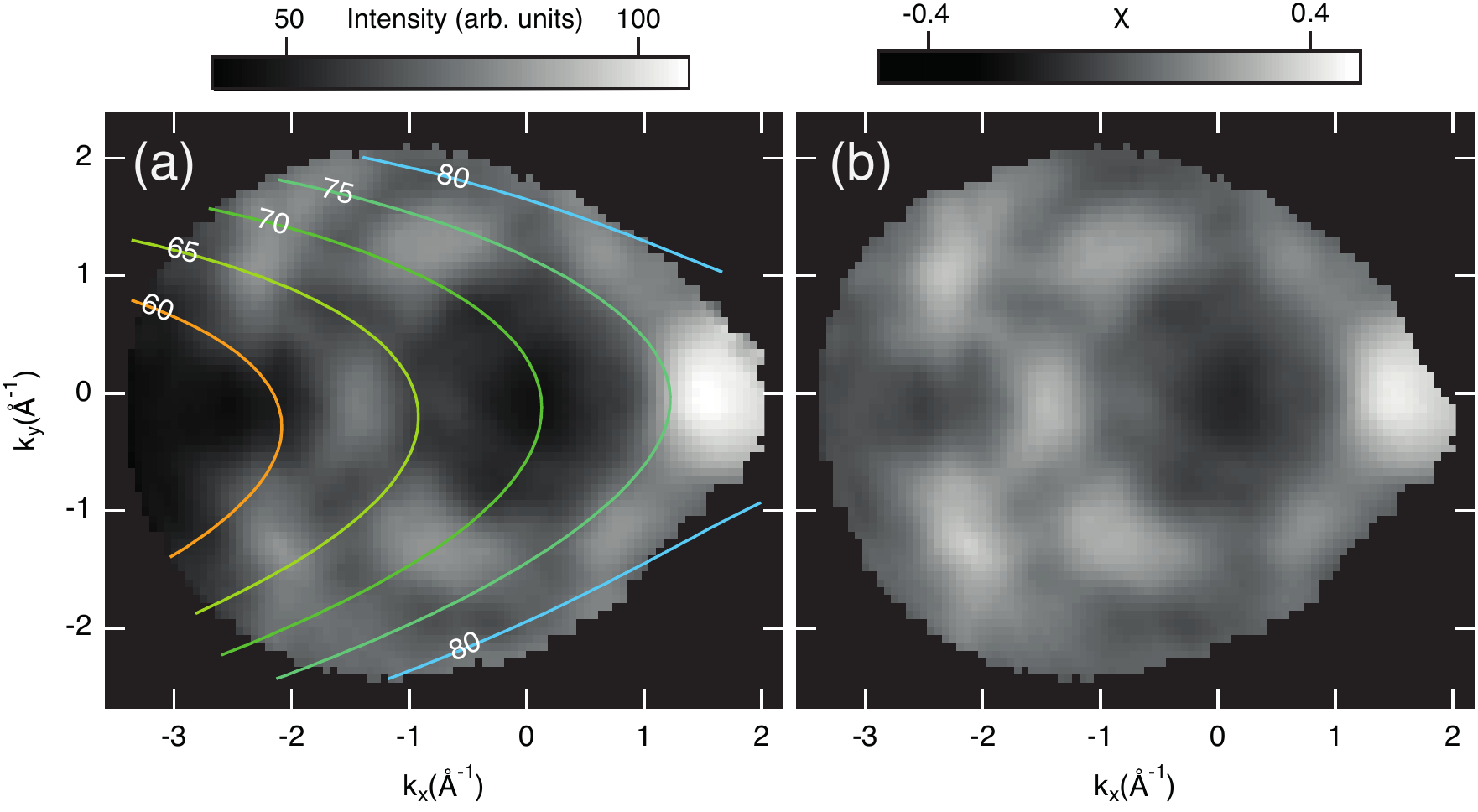}\\
  \caption{
    (a) Photoemission intensity from the \ce{Se}~3d core level, displayed as lateral momentum-resolved peak area.
    The superimposed contours show the 2-dimensional polynomial of degree 2 used to generate the smooth background $I_0(\mathbf{k})$ needed to calculate the modulation function.
    (b) Resulting modulation function.
    }
  \label{fig:S6}
\end{figure}

The modulation function $\chi$ described in the main text has been obtained using a 2-dimensional polynomial of degree 2 fitted to the photoelectron intensity modulation pattern $I_0({\bf k})$.
This is illustrated in Fig.~\ref{fig:S6}(a), and the process yields the modulation function shown in Fig.~\ref{fig:S6}(b).

\subsection{Multiple-scattering simulations}

Simulations of the photoelectron diffraction patterns were performed using the Electron Diffraction in Atomic Cluster package~\cite{Garcia-de-Abajo:2001aa}.
The code uses an input of an atomic cluster surrounding the emitter and several non-structural parameters, from which the diffraction pattern is calculated using multiple scattering theory.
For the simulations, the structural and non-structural parameters were optimized to obtain the best $R$-factor in comparison to the time-integrated XPD pattern.
The structural parameters were fixed to the literature values~\cite{Kuznetsov:2015ab} and only $d_1$, $d_2$ were allowed to vary.
To model the experimental configuration, a fixed cluster was used with an incidence angle of 68$^{\circ}$ for the horizontal linearly polarized light.
The Debye temperature was held fixed at the literature value of 185~K~\cite{Madelung:2004aa}.
The inelastic mean free path was determined to be 5.5~{\AA}, which corresponds well to the value of 4.94~{\AA} that the Tanuma Powell and Penn algorithm~\cite{Tanuma:1994aa} yields at a kinetic energy of 60~eV.
The inner potential V$_0$ was found to be 3.2~eV.

\subsection{Phonon model}

The vibrational frequencies for the one-dimensional infinite chain model in Figure 3 of the main paper were calculated using the following force constants: $\gamma_1=$11.8~Nm$^{-1}$, $\gamma_2=$79.0~Nm$^{-1}$, $\gamma_3=$63.2~Nm$^{-1}$.
These values were chosen such that the resulting dispersion is similar to the dispersion along the bulk $\Gamma$-Z direction calculated using a pseudo-charge model in Ref.~\cite{Zhu:2011ab} (see Fig.~6 in the supplementary material of that paper).
The resulting frequency of the A$^{1}_{1g}$ mode in the model is 2.15~THz.
In order to realize a softer surface (end)-localized mode, such as reported in Refs.~\cite{Sobota:2014aa,Ruckhofer:2020vg}, the surface force constant $\gamma_s$ was chosen to be 10\% smaller than $\gamma_2$.
The frequency of the resulting end mode is 1.77~THz.
For the A$^{1}_{1g}$ mode, the vector describing the relative displacement of the five atoms in the unit cell is $(-0.37,-0.6,0.0,0.6,0.37)$.
For the end mode, the vector for the first quintuple layer is $(-0.14,-0.19,0,0.18,0.14)$ , see inset to Fig~3(b).
For the creation of Figure~3(c) and (d) in the main text, the end mode vibrational amplitude was chosen to be 2.3 times higher than the bulk vibrational amplitude and the two modes are 70~fs out of phase.

%\bibliographystyle{apsrev}
%\bibliography{cophoxpd,local}

\begin{thebibliography}{45}
\expandafter\ifx\csname natexlab\endcsname\relax\def\natexlab#1{#1}\fi
\expandafter\ifx\csname bibnamefont\endcsname\relax
  \def\bibnamefont#1{#1}\fi
\expandafter\ifx\csname bibfnamefont\endcsname\relax
  \def\bibfnamefont#1{#1}\fi
\expandafter\ifx\csname citenamefont\endcsname\relax
  \def\citenamefont#1{#1}\fi
\expandafter\ifx\csname url\endcsname\relax
  \def\url#1{\texttt{#1}}\fi
\expandafter\ifx\csname urlprefix\endcsname\relax\def\urlprefix{URL }\fi
\providecommand{\bibinfo}[2]{#2}
\providecommand{\eprint}[2][]{\url{#2}}

\bibitem[{\citenamefont{Kirilyuk et~al.}(2010)\citenamefont{Kirilyuk, Kimel,
  and Rasing}}]{Kirilyuk:2010vd}
\bibinfo{author}{\bibfnamefont{A.}~\bibnamefont{Kirilyuk}},
  \bibinfo{author}{\bibfnamefont{A.~V.} \bibnamefont{Kimel}}, \bibnamefont{and}
  \bibinfo{author}{\bibfnamefont{T.}~\bibnamefont{Rasing}},
  \bibinfo{journal}{Rev. Mod. Phys.} \textbf{\bibinfo{volume}{82}},
  \bibinfo{pages}{2731} (\bibinfo{year}{2010}).

\bibitem[{\citenamefont{Giannetti et~al.}(2016)\citenamefont{Giannetti, Capone,
  Fausti, Fabrizio, Parmigiani, and Mihailovic}}]{Giannetti:2016uf}
\bibinfo{author}{\bibfnamefont{C.}~\bibnamefont{Giannetti}},
  \bibinfo{author}{\bibfnamefont{M.}~\bibnamefont{Capone}},
  \bibinfo{author}{\bibfnamefont{D.}~\bibnamefont{Fausti}},
  \bibinfo{author}{\bibfnamefont{M.}~\bibnamefont{Fabrizio}},
  \bibinfo{author}{\bibfnamefont{F.}~\bibnamefont{Parmigiani}},
  \bibnamefont{and}
  \bibinfo{author}{\bibfnamefont{D.}~\bibnamefont{Mihailovic}},
  \bibinfo{journal}{Advances in Physics} \textbf{\bibinfo{volume}{65}},
  \bibinfo{pages}{58} (\bibinfo{year}{2016}).

\bibitem[{\citenamefont{Buzzi et~al.}(2018)\citenamefont{Buzzi, F{\"o}rst,
  Mankowsky, and Cavalleri}}]{Buzzi:2018wu}
\bibinfo{author}{\bibfnamefont{M.}~\bibnamefont{Buzzi}},
  \bibinfo{author}{\bibfnamefont{M.}~\bibnamefont{F{\"o}rst}},
  \bibinfo{author}{\bibfnamefont{R.}~\bibnamefont{Mankowsky}},
  \bibnamefont{and}
  \bibinfo{author}{\bibfnamefont{A.}~\bibnamefont{Cavalleri}},
  \bibinfo{journal}{Nature Reviews Materials} \textbf{\bibinfo{volume}{3}},
  \bibinfo{pages}{299} (\bibinfo{year}{2018}).

\bibitem[{\citenamefont{de~la Torre et~al.}(2021)\citenamefont{de~la Torre,
  Kennes, Claassen, Gerber, McIver, and Sentef}}]{Torre:2021wg}
\bibinfo{author}{\bibfnamefont{A.}~\bibnamefont{de~la Torre}},
  \bibinfo{author}{\bibfnamefont{D.~M.} \bibnamefont{Kennes}},
  \bibinfo{author}{\bibfnamefont{M.}~\bibnamefont{Claassen}},
  \bibinfo{author}{\bibfnamefont{S.}~\bibnamefont{Gerber}},
  \bibinfo{author}{\bibfnamefont{J.~W.} \bibnamefont{McIver}},
  \bibnamefont{and} \bibinfo{author}{\bibfnamefont{M.~A.}
  \bibnamefont{Sentef}}, \bibinfo{journal}{Rev. Mod. Phys.}
  \textbf{\bibinfo{volume}{93}}, \bibinfo{pages}{041002}
  (\bibinfo{year}{2021}).

\bibitem[{\citenamefont{Maiuri et~al.}(2020)\citenamefont{Maiuri, Garavelli,
  and Cerullo}}]{Maiuri:2020ve}
\bibinfo{author}{\bibfnamefont{M.}~\bibnamefont{Maiuri}},
  \bibinfo{author}{\bibfnamefont{M.}~\bibnamefont{Garavelli}},
  \bibnamefont{and} \bibinfo{author}{\bibfnamefont{G.}~\bibnamefont{Cerullo}},
  \bibinfo{journal}{Journal of the American Chemical Society}
  \textbf{\bibinfo{volume}{142}}, \bibinfo{pages}{3} (\bibinfo{year}{2020}).

\bibitem[{\citenamefont{Bovensiepen and Kirchmann}(2012)}]{Bovensiepen:2012tm}
\bibinfo{author}{\bibfnamefont{U.}~\bibnamefont{Bovensiepen}} \bibnamefont{and}
  \bibinfo{author}{\bibfnamefont{P.~S.} \bibnamefont{Kirchmann}},
  \bibinfo{journal}{Laser \& Photonics Reviews} \textbf{\bibinfo{volume}{6}},
  \bibinfo{pages}{589} (\bibinfo{year}{2012}).

\bibitem[{\citenamefont{Lindenberg et~al.}(2017)\citenamefont{Lindenberg,
  Johnson, and Reis}}]{Lindenberg:2017aa}
\bibinfo{author}{\bibfnamefont{A.~M.} \bibnamefont{Lindenberg}},
  \bibinfo{author}{\bibfnamefont{S.~L.} \bibnamefont{Johnson}},
  \bibnamefont{and} \bibinfo{author}{\bibfnamefont{D.~A.} \bibnamefont{Reis}},
  \bibinfo{journal}{Annual Review of Materials Research}
  \textbf{\bibinfo{volume}{47}}, \bibinfo{pages}{425} (\bibinfo{year}{2017}).

\bibitem[{\citenamefont{Gerber et~al.}(2017)\citenamefont{Gerber, Yang, Zhu,
  Soifer, Sobota, Rebec, Lee, Jia, Moritz, Jia et~al.}}]{Gerber:2017aa}
\bibinfo{author}{\bibfnamefont{S.}~\bibnamefont{Gerber}},
  \bibinfo{author}{\bibfnamefont{S.-L.} \bibnamefont{Yang}},
  \bibinfo{author}{\bibfnamefont{D.}~\bibnamefont{Zhu}},
  \bibinfo{author}{\bibfnamefont{H.}~\bibnamefont{Soifer}},
  \bibinfo{author}{\bibfnamefont{J.~A.} \bibnamefont{Sobota}},
  \bibinfo{author}{\bibfnamefont{S.}~\bibnamefont{Rebec}},
  \bibinfo{author}{\bibfnamefont{J.~J.} \bibnamefont{Lee}},
  \bibinfo{author}{\bibfnamefont{T.}~\bibnamefont{Jia}},
  \bibinfo{author}{\bibfnamefont{B.}~\bibnamefont{Moritz}},
  \bibinfo{author}{\bibfnamefont{C.}~\bibnamefont{Jia}}, \bibnamefont{et~al.},
  \bibinfo{journal}{Science} \textbf{\bibinfo{volume}{357}},
  \bibinfo{pages}{71} (\bibinfo{year}{2017}).

\bibitem[{\citenamefont{Giovannini et~al.}(2020)\citenamefont{Giovannini,
  H{\"u}bener, Sato, and Rubio}}]{Giovannini:2020aa}
\bibinfo{author}{\bibfnamefont{U.~D.} \bibnamefont{Giovannini}},
  \bibinfo{author}{\bibfnamefont{H.}~\bibnamefont{H{\"u}bener}},
  \bibinfo{author}{\bibfnamefont{S.~A.} \bibnamefont{Sato}}, \bibnamefont{and}
  \bibinfo{author}{\bibfnamefont{A.}~\bibnamefont{Rubio}},
  \bibinfo{journal}{Physical Review Letters} \textbf{\bibinfo{volume}{125}},
  \bibinfo{pages}{136401} (\bibinfo{year}{2020}).

\bibitem[{\citenamefont{Woodruff}(1992)}]{Woodruff:1992aa}
\bibinfo{author}{\bibfnamefont{D.~P.} \bibnamefont{Woodruff}}, in
  \emph{\bibinfo{booktitle}{Angle-resolved photoemission}}, edited by
  \bibinfo{editor}{\bibfnamefont{S.~D.} \bibnamefont{Kevan}}
  (\bibinfo{publisher}{Elsevier}, \bibinfo{address}{Amsterdam},
  \bibinfo{year}{1992}).

\bibitem[{\citenamefont{Woodruff and Bradshaw}(1994)}]{Woodruff:1994ab}
\bibinfo{author}{\bibfnamefont{D.~P.} \bibnamefont{Woodruff}} \bibnamefont{and}
  \bibinfo{author}{\bibfnamefont{A.~M.} \bibnamefont{Bradshaw}},
  \bibinfo{journal}{Reports on Progress in Physics}
  \textbf{\bibinfo{volume}{57}}, \bibinfo{pages}{1029} (\bibinfo{year}{1994}).

\bibitem[{\citenamefont{Osterwalder et~al.}(1995)\citenamefont{Osterwalder,
  Aebi, Fasel, Naumovic, Schwaller, Kreutz, Schlapbach, Abukawa, and
  Kono}}]{Osterwalder:1995vf}
\bibinfo{author}{\bibfnamefont{J.}~\bibnamefont{Osterwalder}},
  \bibinfo{author}{\bibfnamefont{P.}~\bibnamefont{Aebi}},
  \bibinfo{author}{\bibfnamefont{R.}~\bibnamefont{Fasel}},
  \bibinfo{author}{\bibfnamefont{D.}~\bibnamefont{Naumovic}},
  \bibinfo{author}{\bibfnamefont{P.}~\bibnamefont{Schwaller}},
  \bibinfo{author}{\bibfnamefont{T.}~\bibnamefont{Kreutz}},
  \bibinfo{author}{\bibfnamefont{L.}~\bibnamefont{Schlapbach}},
  \bibinfo{author}{\bibfnamefont{T.}~\bibnamefont{Abukawa}}, \bibnamefont{and}
  \bibinfo{author}{\bibfnamefont{S.}~\bibnamefont{Kono}},
  \bibinfo{journal}{Surface Science} \textbf{\bibinfo{volume}{331-333}},
  \bibinfo{pages}{1002} (\bibinfo{year}{1995}).

\bibitem[{\citenamefont{Hofmann et~al.}(1994)\citenamefont{Hofmann, Schindler,
  Bao, Bradshaw, and Woodruff}}]{Hofmann:1994ae}
\bibinfo{author}{\bibfnamefont{P.}~\bibnamefont{Hofmann}},
  \bibinfo{author}{\bibfnamefont{K.-M.} \bibnamefont{Schindler}},
  \bibinfo{author}{\bibfnamefont{S.}~\bibnamefont{Bao}},
  \bibinfo{author}{\bibfnamefont{A.~M.} \bibnamefont{Bradshaw}},
  \bibnamefont{and} \bibinfo{author}{\bibfnamefont{D.~P.}
  \bibnamefont{Woodruff}}, \bibinfo{journal}{Nature}
  \textbf{\bibinfo{volume}{368}}, \bibinfo{pages}{131} (\bibinfo{year}{1994}).

\bibitem[{\citenamefont{Kuznetsov et~al.}(2015)\citenamefont{Kuznetsov,
  Yashina, S\'anchez-Barriga, Ogorodnikov, Vorokh, Volykhov, Koch, Neudachina,
  Tamm, Sirotina et~al.}}]{Kuznetsov:2015ab}
\bibinfo{author}{\bibfnamefont{M.~V.} \bibnamefont{Kuznetsov}},
  \bibinfo{author}{\bibfnamefont{L.~V.} \bibnamefont{Yashina}},
  \bibinfo{author}{\bibfnamefont{J.}~\bibnamefont{S\'anchez-Barriga}},
  \bibinfo{author}{\bibfnamefont{I.~I.} \bibnamefont{Ogorodnikov}},
  \bibinfo{author}{\bibfnamefont{A.~S.} \bibnamefont{Vorokh}},
  \bibinfo{author}{\bibfnamefont{A.~A.} \bibnamefont{Volykhov}},
  \bibinfo{author}{\bibfnamefont{R.~J.} \bibnamefont{Koch}},
  \bibinfo{author}{\bibfnamefont{V.~S.} \bibnamefont{Neudachina}},
  \bibinfo{author}{\bibfnamefont{M.~E.} \bibnamefont{Tamm}},
  \bibinfo{author}{\bibfnamefont{A.~P.} \bibnamefont{Sirotina}},
  \bibnamefont{et~al.}, \bibinfo{journal}{Phys. Rev. B}
  \textbf{\bibinfo{volume}{91}}, \bibinfo{pages}{085402}
  (\bibinfo{year}{2015}).

\bibitem[{\citenamefont{Bao et~al.}(1994)\citenamefont{Bao, Hofmann, Schindler,
  Fritzsche, Bradshaw, Woodruff, Casado, and Asensio}}]{BAO:1994aa}
\bibinfo{author}{\bibfnamefont{S.}~\bibnamefont{Bao}},
  \bibinfo{author}{\bibfnamefont{P.}~\bibnamefont{Hofmann}},
  \bibinfo{author}{\bibfnamefont{K.~M.} \bibnamefont{Schindler}},
  \bibinfo{author}{\bibfnamefont{V.}~\bibnamefont{Fritzsche}},
  \bibinfo{author}{\bibfnamefont{A.~M.} \bibnamefont{Bradshaw}},
  \bibinfo{author}{\bibfnamefont{D.~P.} \bibnamefont{Woodruff}},
  \bibinfo{author}{\bibfnamefont{C.}~\bibnamefont{Casado}}, \bibnamefont{and}
  \bibinfo{author}{\bibfnamefont{M.~C.} \bibnamefont{Asensio}},
  \bibinfo{journal}{Journal of Physics-Condensed Matter}
  \textbf{\bibinfo{volume}{6}}, \bibinfo{pages}{L93} (\bibinfo{year}{1994}).

\bibitem[{\citenamefont{Boll et~al.}(2014)\citenamefont{Boll, Rouz{\'e}e,
  Adolph, Anielski, Aquila, Bari, Bomme, Bostedt, Bozek, Chapman
  et~al.}}]{Boll:2014up}
\bibinfo{author}{\bibfnamefont{R.}~\bibnamefont{Boll}},
  \bibinfo{author}{\bibfnamefont{A.}~\bibnamefont{Rouz{\'e}e}},
  \bibinfo{author}{\bibfnamefont{M.}~\bibnamefont{Adolph}},
  \bibinfo{author}{\bibfnamefont{D.}~\bibnamefont{Anielski}},
  \bibinfo{author}{\bibfnamefont{A.}~\bibnamefont{Aquila}},
  \bibinfo{author}{\bibfnamefont{S.}~\bibnamefont{Bari}},
  \bibinfo{author}{\bibfnamefont{C.}~\bibnamefont{Bomme}},
  \bibinfo{author}{\bibfnamefont{C.}~\bibnamefont{Bostedt}},
  \bibinfo{author}{\bibfnamefont{J.~D.} \bibnamefont{Bozek}},
  \bibinfo{author}{\bibfnamefont{H.~N.} \bibnamefont{Chapman}},
  \bibnamefont{et~al.}, \bibinfo{journal}{Faraday Discuss.}
  \textbf{\bibinfo{volume}{171}}, \bibinfo{pages}{57} (\bibinfo{year}{2014}).

\bibitem[{\citenamefont{Kastirke et~al.}(2020)\citenamefont{Kastirke,
  Sch{\"o}ffler, Weller, Rist, Boll, Anders, Baumann, Eckart, Erk, De~Fanis
  et~al.}}]{Kastirke:2020aa}
\bibinfo{author}{\bibfnamefont{G.}~\bibnamefont{Kastirke}},
  \bibinfo{author}{\bibfnamefont{M.~S.} \bibnamefont{Sch{\"o}ffler}},
  \bibinfo{author}{\bibfnamefont{M.}~\bibnamefont{Weller}},
  \bibinfo{author}{\bibfnamefont{J.}~\bibnamefont{Rist}},
  \bibinfo{author}{\bibfnamefont{R.}~\bibnamefont{Boll}},
  \bibinfo{author}{\bibfnamefont{N.}~\bibnamefont{Anders}},
  \bibinfo{author}{\bibfnamefont{T.~M.} \bibnamefont{Baumann}},
  \bibinfo{author}{\bibfnamefont{S.}~\bibnamefont{Eckart}},
  \bibinfo{author}{\bibfnamefont{B.}~\bibnamefont{Erk}},
  \bibinfo{author}{\bibfnamefont{A.}~\bibnamefont{De~Fanis}},
  \bibnamefont{et~al.}, \bibinfo{journal}{Physical Review X}
  \textbf{\bibinfo{volume}{10}}, \bibinfo{pages}{021052}
  (\bibinfo{year}{2020}).

\bibitem[{\citenamefont{Greif et~al.}(2016)\citenamefont{Greif, Kasmi,
  Castiglioni, Lucchini, Gallmann, Keller, Osterwalder, and
  Hengsberger}}]{Greif:2016aa}
\bibinfo{author}{\bibfnamefont{M.}~\bibnamefont{Greif}},
  \bibinfo{author}{\bibfnamefont{L.}~\bibnamefont{Kasmi}},
  \bibinfo{author}{\bibfnamefont{L.}~\bibnamefont{Castiglioni}},
  \bibinfo{author}{\bibfnamefont{M.}~\bibnamefont{Lucchini}},
  \bibinfo{author}{\bibfnamefont{L.}~\bibnamefont{Gallmann}},
  \bibinfo{author}{\bibfnamefont{U.}~\bibnamefont{Keller}},
  \bibinfo{author}{\bibfnamefont{J.}~\bibnamefont{Osterwalder}},
  \bibnamefont{and}
  \bibinfo{author}{\bibfnamefont{M.}~\bibnamefont{Hengsberger}},
  \bibinfo{journal}{Phys. Rev. B} \textbf{\bibinfo{volume}{94}},
  \bibinfo{pages}{054309} (\bibinfo{year}{2016}).

\bibitem[{\citenamefont{Ang et~al.}(2020)\citenamefont{Ang, Fukatsu, Kimura,
  Yamamoto, Yonezawa, Nitta, Fleurence, Yamamoto, Matsuda, Yamada-Takamura
  et~al.}}]{Ang:2020aa}
\bibinfo{author}{\bibfnamefont{A.~K.~R.} \bibnamefont{Ang}},
  \bibinfo{author}{\bibfnamefont{Y.}~\bibnamefont{Fukatsu}},
  \bibinfo{author}{\bibfnamefont{K.}~\bibnamefont{Kimura}},
  \bibinfo{author}{\bibfnamefont{Y.}~\bibnamefont{Yamamoto}},
  \bibinfo{author}{\bibfnamefont{T.}~\bibnamefont{Yonezawa}},
  \bibinfo{author}{\bibfnamefont{H.}~\bibnamefont{Nitta}},
  \bibinfo{author}{\bibfnamefont{A.}~\bibnamefont{Fleurence}},
  \bibinfo{author}{\bibfnamefont{S.}~\bibnamefont{Yamamoto}},
  \bibinfo{author}{\bibfnamefont{I.}~\bibnamefont{Matsuda}},
  \bibinfo{author}{\bibfnamefont{Y.}~\bibnamefont{Yamada-Takamura}},
  \bibnamefont{et~al.}, \bibinfo{journal}{Japanese Journal of Applied Physics}
  \textbf{\bibinfo{volume}{59}}, \bibinfo{pages}{100902}
  (\bibinfo{year}{2020}).

\bibitem[{\citenamefont{Qi et~al.}(2010)\citenamefont{Qi, Chen, Yu,
  Cadden-Zimansky, Smirnov, Tolk, Miotkowski, Cao, Chen, Wu
  et~al.}}]{Qi:2010ur}
\bibinfo{author}{\bibfnamefont{J.}~\bibnamefont{Qi}},
  \bibinfo{author}{\bibfnamefont{X.}~\bibnamefont{Chen}},
  \bibinfo{author}{\bibfnamefont{W.}~\bibnamefont{Yu}},
  \bibinfo{author}{\bibfnamefont{P.}~\bibnamefont{Cadden-Zimansky}},
  \bibinfo{author}{\bibfnamefont{D.}~\bibnamefont{Smirnov}},
  \bibinfo{author}{\bibfnamefont{N.~H.} \bibnamefont{Tolk}},
  \bibinfo{author}{\bibfnamefont{I.}~\bibnamefont{Miotkowski}},
  \bibinfo{author}{\bibfnamefont{H.}~\bibnamefont{Cao}},
  \bibinfo{author}{\bibfnamefont{Y.~P.} \bibnamefont{Chen}},
  \bibinfo{author}{\bibfnamefont{Y.}~\bibnamefont{Wu}}, \bibnamefont{et~al.},
  \bibinfo{journal}{Applied Physics Letters} \textbf{\bibinfo{volume}{97}},
  \bibinfo{pages}{182102} (\bibinfo{year}{2010}).

\bibitem[{\citenamefont{Kumar et~al.}(2011)\citenamefont{Kumar, Ruzicka, Butch,
  Syers, Kirshenbaum, Paglione, and Zhao}}]{Kumar:2011ue}
\bibinfo{author}{\bibfnamefont{N.}~\bibnamefont{Kumar}},
  \bibinfo{author}{\bibfnamefont{B.~A.} \bibnamefont{Ruzicka}},
  \bibinfo{author}{\bibfnamefont{N.~P.} \bibnamefont{Butch}},
  \bibinfo{author}{\bibfnamefont{P.}~\bibnamefont{Syers}},
  \bibinfo{author}{\bibfnamefont{K.}~\bibnamefont{Kirshenbaum}},
  \bibinfo{author}{\bibfnamefont{J.}~\bibnamefont{Paglione}}, \bibnamefont{and}
  \bibinfo{author}{\bibfnamefont{H.}~\bibnamefont{Zhao}},
  \bibinfo{journal}{Phys. Rev. B} \textbf{\bibinfo{volume}{83}},
  \bibinfo{pages}{235306} (\bibinfo{year}{2011}).

\bibitem[{\citenamefont{Hu et~al.}(2018)\citenamefont{Hu, Igarashi, Sasagawa,
  Nakamura, and Misochko}}]{Hu:2018ab}
\bibinfo{author}{\bibfnamefont{J.}~\bibnamefont{Hu}},
  \bibinfo{author}{\bibfnamefont{K.}~\bibnamefont{Igarashi}},
  \bibinfo{author}{\bibfnamefont{T.}~\bibnamefont{Sasagawa}},
  \bibinfo{author}{\bibfnamefont{K.~G.} \bibnamefont{Nakamura}},
  \bibnamefont{and} \bibinfo{author}{\bibfnamefont{O.~V.}
  \bibnamefont{Misochko}}, \bibinfo{journal}{Applied Physics Letters}
  \textbf{\bibinfo{volume}{112}}, \bibinfo{pages}{031901}
  (\bibinfo{year}{2018}).

\bibitem[{\citenamefont{Sobota et~al.}(2014)\citenamefont{Sobota, Yang,
  Leuenberger, Kemper, Analytis, Fisher, Kirchmann, Devereaux, and
  Shen}}]{Sobota:2014aa}
\bibinfo{author}{\bibfnamefont{J.~A.} \bibnamefont{Sobota}},
  \bibinfo{author}{\bibfnamefont{S.-L.} \bibnamefont{Yang}},
  \bibinfo{author}{\bibfnamefont{D.}~\bibnamefont{Leuenberger}},
  \bibinfo{author}{\bibfnamefont{A.~F.} \bibnamefont{Kemper}},
  \bibinfo{author}{\bibfnamefont{J.~G.} \bibnamefont{Analytis}},
  \bibinfo{author}{\bibfnamefont{I.~R.} \bibnamefont{Fisher}},
  \bibinfo{author}{\bibfnamefont{P.~S.} \bibnamefont{Kirchmann}},
  \bibinfo{author}{\bibfnamefont{T.~P.} \bibnamefont{Devereaux}},
  \bibnamefont{and} \bibinfo{author}{\bibfnamefont{Z.-X.} \bibnamefont{Shen}},
  \bibinfo{journal}{Phys. Rev. Lett.} \textbf{\bibinfo{volume}{113}},
  \bibinfo{pages}{157401} (\bibinfo{year}{2014}).

\bibitem[{\citenamefont{Ruckhofer et~al.}(2020)\citenamefont{Ruckhofer, Campi,
  Bremholm, Hofmann, Benedek, Bernasconi, Ernst, and
  Tamt\"ogl}}]{Ruckhofer:2020vg}
\bibinfo{author}{\bibfnamefont{A.}~\bibnamefont{Ruckhofer}},
  \bibinfo{author}{\bibfnamefont{D.}~\bibnamefont{Campi}},
  \bibinfo{author}{\bibfnamefont{M.}~\bibnamefont{Bremholm}},
  \bibinfo{author}{\bibfnamefont{P.}~\bibnamefont{Hofmann}},
  \bibinfo{author}{\bibfnamefont{G.}~\bibnamefont{Benedek}},
  \bibinfo{author}{\bibfnamefont{M.}~\bibnamefont{Bernasconi}},
  \bibinfo{author}{\bibfnamefont{W.~E.} \bibnamefont{Ernst}}, \bibnamefont{and}
  \bibinfo{author}{\bibfnamefont{A.}~\bibnamefont{Tamt\"ogl}},
  \bibinfo{journal}{Phys. Rev. Research} \textbf{\bibinfo{volume}{2}},
  \bibinfo{pages}{023186} (\bibinfo{year}{2020}).

\bibitem[{\citenamefont{Kutnyakhov et~al.}(2020)\citenamefont{Kutnyakhov, Xian,
  Dendzik, Heber, Pressacco, Agustsson, Wenthaus, Meyer, Gieschen, Mercurio
  et~al.}}]{Kutnyakhov:2020aa}
\bibinfo{author}{\bibfnamefont{D.}~\bibnamefont{Kutnyakhov}},
  \bibinfo{author}{\bibfnamefont{R.~P.} \bibnamefont{Xian}},
  \bibinfo{author}{\bibfnamefont{M.}~\bibnamefont{Dendzik}},
  \bibinfo{author}{\bibfnamefont{M.}~\bibnamefont{Heber}},
  \bibinfo{author}{\bibfnamefont{F.}~\bibnamefont{Pressacco}},
  \bibinfo{author}{\bibfnamefont{S.~Y.} \bibnamefont{Agustsson}},
  \bibinfo{author}{\bibfnamefont{L.}~\bibnamefont{Wenthaus}},
  \bibinfo{author}{\bibfnamefont{H.}~\bibnamefont{Meyer}},
  \bibinfo{author}{\bibfnamefont{S.}~\bibnamefont{Gieschen}},
  \bibinfo{author}{\bibfnamefont{G.}~\bibnamefont{Mercurio}},
  \bibnamefont{et~al.}, \bibinfo{journal}{Review of Scientific Instruments}
  \textbf{\bibinfo{volume}{91}}, \bibinfo{pages}{013109}
  (\bibinfo{year}{2020}).

\bibitem[{\citenamefont{Gerasimova et~al.}(2011)\citenamefont{Gerasimova,
  Dziarzhytski, and Feldhaus}}]{Gerasimova:2011aa}
\bibinfo{author}{\bibfnamefont{N.}~\bibnamefont{Gerasimova}},
  \bibinfo{author}{\bibfnamefont{S.}~\bibnamefont{Dziarzhytski}},
  \bibnamefont{and} \bibinfo{author}{\bibfnamefont{J.}~\bibnamefont{Feldhaus}},
  \bibinfo{journal}{Journal of Modern Optics} \textbf{\bibinfo{volume}{58}},
  \bibinfo{pages}{1480} (\bibinfo{year}{2011}).

\bibitem[{\citenamefont{Doniach and Sunjic}(1970)}]{Doniach:1970aa}
\bibinfo{author}{\bibfnamefont{S.}~\bibnamefont{Doniach}} \bibnamefont{and}
  \bibinfo{author}{\bibfnamefont{M.}~\bibnamefont{Sunjic}},
  \bibinfo{journal}{Journal of Physics C: Solid State Physics}
  \textbf{\bibinfo{volume}{3}}, \bibinfo{pages}{285 } (\bibinfo{year}{1970}).

\bibitem[{\citenamefont{Dendzik et~al.}(2020)\citenamefont{Dendzik, Xian,
  Perfetto, Sangalli, Kutnyakhov, Dong, Beaulieu, Pincelli, Pressacco, Curcio
  et~al.}}]{Dendzik:2020aa}
\bibinfo{author}{\bibfnamefont{M.}~\bibnamefont{Dendzik}},
  \bibinfo{author}{\bibfnamefont{R.~P.} \bibnamefont{Xian}},
  \bibinfo{author}{\bibfnamefont{E.}~\bibnamefont{Perfetto}},
  \bibinfo{author}{\bibfnamefont{D.}~\bibnamefont{Sangalli}},
  \bibinfo{author}{\bibfnamefont{D.}~\bibnamefont{Kutnyakhov}},
  \bibinfo{author}{\bibfnamefont{S.}~\bibnamefont{Dong}},
  \bibinfo{author}{\bibfnamefont{S.}~\bibnamefont{Beaulieu}},
  \bibinfo{author}{\bibfnamefont{T.}~\bibnamefont{Pincelli}},
  \bibinfo{author}{\bibfnamefont{F.}~\bibnamefont{Pressacco}},
  \bibinfo{author}{\bibfnamefont{D.}~\bibnamefont{Curcio}},
  \bibnamefont{et~al.}, \bibinfo{journal}{Phys. Rev. Lett.}
  \textbf{\bibinfo{volume}{125}}, \bibinfo{pages}{096401}
  (\bibinfo{year}{2020}).

\bibitem[{\citenamefont{Curcio et~al.}(2021)\citenamefont{Curcio, Pakdel,
  Volckaert, Miwa, Ulstrup, Lanat\`a, Bianchi, Kutnyakhov, Pressacco, Brenner
  et~al.}}]{Curcio:2021uh}
\bibinfo{author}{\bibfnamefont{D.}~\bibnamefont{Curcio}},
  \bibinfo{author}{\bibfnamefont{S.}~\bibnamefont{Pakdel}},
  \bibinfo{author}{\bibfnamefont{K.}~\bibnamefont{Volckaert}},
  \bibinfo{author}{\bibfnamefont{J.~A.} \bibnamefont{Miwa}},
  \bibinfo{author}{\bibfnamefont{S.}~\bibnamefont{Ulstrup}},
  \bibinfo{author}{\bibfnamefont{N.}~\bibnamefont{Lanat\`a}},
  \bibinfo{author}{\bibfnamefont{M.}~\bibnamefont{Bianchi}},
  \bibinfo{author}{\bibfnamefont{D.}~\bibnamefont{Kutnyakhov}},
  \bibinfo{author}{\bibfnamefont{F.}~\bibnamefont{Pressacco}},
  \bibinfo{author}{\bibfnamefont{G.}~\bibnamefont{Brenner}},
  \bibnamefont{et~al.}, \bibinfo{journal}{Phys. Rev. B}
  \textbf{\bibinfo{volume}{104}}, \bibinfo{pages}{L161104}
  (\bibinfo{year}{2021}).

\bibitem[{\citenamefont{Garc\'ia~de Abajo
  et~al.}(2001)\citenamefont{Garc\'ia~de Abajo, Van~Hove, and
  Fadley}}]{Garcia-de-Abajo:2001aa}
\bibinfo{author}{\bibfnamefont{F.~J.} \bibnamefont{Garc\'ia~de Abajo}},
  \bibinfo{author}{\bibfnamefont{M.~A.} \bibnamefont{Van~Hove}},
  \bibnamefont{and} \bibinfo{author}{\bibfnamefont{C.~S.}
  \bibnamefont{Fadley}}, \bibinfo{journal}{Phys. Rev. B}
  \textbf{\bibinfo{volume}{63}}, \bibinfo{pages}{075404}
  (\bibinfo{year}{2001}).

\bibitem[{\citenamefont{Nakajima}(1963)}]{Nakajima:1963aa}
\bibinfo{author}{\bibfnamefont{S.}~\bibnamefont{Nakajima}},
  \bibinfo{journal}{J. Phys. Chem. Solids} \textbf{\bibinfo{volume}{24}},
  \bibinfo{pages}{479} (\bibinfo{year}{1963}).

\bibitem[{\citenamefont{Bana et~al.}(2018)\citenamefont{Bana, Travaglia,
  Bignardi, Lacovig, Sanders, Dendzik, Michiardi, Bianchi, Lizzit, Presel
  et~al.}}]{Bana:2018aa}
\bibinfo{author}{\bibfnamefont{H.}~\bibnamefont{Bana}},
  \bibinfo{author}{\bibfnamefont{E.}~\bibnamefont{Travaglia}},
  \bibinfo{author}{\bibfnamefont{L.}~\bibnamefont{Bignardi}},
  \bibinfo{author}{\bibfnamefont{P.}~\bibnamefont{Lacovig}},
  \bibinfo{author}{\bibfnamefont{C.~E.} \bibnamefont{Sanders}},
  \bibinfo{author}{\bibfnamefont{M.}~\bibnamefont{Dendzik}},
  \bibinfo{author}{\bibfnamefont{M.}~\bibnamefont{Michiardi}},
  \bibinfo{author}{\bibfnamefont{M.}~\bibnamefont{Bianchi}},
  \bibinfo{author}{\bibfnamefont{D.}~\bibnamefont{Lizzit}},
  \bibinfo{author}{\bibfnamefont{F.}~\bibnamefont{Presel}},
  \bibnamefont{et~al.}, \bibinfo{journal}{2D Materials}
  \textbf{\bibinfo{volume}{5}}, \bibinfo{pages}{035012} (\bibinfo{year}{2018}).

\bibitem[{\citenamefont{dos Reis et~al.}(2013)\citenamefont{dos Reis, Barreto,
  Bianchi, Ribeiro, Soares, Silva, de~Carvalho, Rawle, Hoesch, Nicklin
  et~al.}}]{Reis:2013aa}
\bibinfo{author}{\bibfnamefont{D.~D.} \bibnamefont{dos Reis}},
  \bibinfo{author}{\bibfnamefont{L.}~\bibnamefont{Barreto}},
  \bibinfo{author}{\bibfnamefont{M.}~\bibnamefont{Bianchi}},
  \bibinfo{author}{\bibfnamefont{G.~A.~S.} \bibnamefont{Ribeiro}},
  \bibinfo{author}{\bibfnamefont{E.~A.} \bibnamefont{Soares}},
  \bibinfo{author}{\bibfnamefont{W.~S.~o.~e.} \bibnamefont{Silva}},
  \bibinfo{author}{\bibfnamefont{V.~E.} \bibnamefont{de~Carvalho}},
  \bibinfo{author}{\bibfnamefont{J.}~\bibnamefont{Rawle}},
  \bibinfo{author}{\bibfnamefont{M.}~\bibnamefont{Hoesch}},
  \bibinfo{author}{\bibfnamefont{C.}~\bibnamefont{Nicklin}},
  \bibnamefont{et~al.}, \bibinfo{journal}{Phys. Rev. B}
  \textbf{\bibinfo{volume}{88}}, \bibinfo{pages}{041404}
  (\bibinfo{year}{2013}).

\bibitem[{\citenamefont{Kim et~al.}(2021)\citenamefont{Kim, Kim, Kim, Choi,
  Yun, Kim, Lim, Kim, Chun, Park et~al.}}]{Kim:2021vf}
\bibinfo{author}{\bibfnamefont{S.}~\bibnamefont{Kim}},
  \bibinfo{author}{\bibfnamefont{Y.}~\bibnamefont{Kim}},
  \bibinfo{author}{\bibfnamefont{J.}~\bibnamefont{Kim}},
  \bibinfo{author}{\bibfnamefont{S.}~\bibnamefont{Choi}},
  \bibinfo{author}{\bibfnamefont{K.}~\bibnamefont{Yun}},
  \bibinfo{author}{\bibfnamefont{D.}~\bibnamefont{Kim}},
  \bibinfo{author}{\bibfnamefont{S.~Y.} \bibnamefont{Lim}},
  \bibinfo{author}{\bibfnamefont{S.}~\bibnamefont{Kim}},
  \bibinfo{author}{\bibfnamefont{S.~H.} \bibnamefont{Chun}},
  \bibinfo{author}{\bibfnamefont{J.}~\bibnamefont{Park}}, \bibnamefont{et~al.},
  \bibinfo{journal}{Nano Letters} \textbf{\bibinfo{volume}{21}},
  \bibinfo{pages}{8554} (\bibinfo{year}{2021}).

\bibitem[{\citenamefont{Johnson et~al.}(2013)\citenamefont{Johnson, Beaud,
  M\"ohr-Vorobeva, Caviezel, Ingold, and Milne}}]{Johnson:2013wc}
\bibinfo{author}{\bibfnamefont{S.~L.} \bibnamefont{Johnson}},
  \bibinfo{author}{\bibfnamefont{P.}~\bibnamefont{Beaud}},
  \bibinfo{author}{\bibfnamefont{E.}~\bibnamefont{M\"ohr-Vorobeva}},
  \bibinfo{author}{\bibfnamefont{A.}~\bibnamefont{Caviezel}},
  \bibinfo{author}{\bibfnamefont{G.}~\bibnamefont{Ingold}}, \bibnamefont{and}
  \bibinfo{author}{\bibfnamefont{C.~J.} \bibnamefont{Milne}},
  \bibinfo{journal}{Phys. Rev. B} \textbf{\bibinfo{volume}{87}},
  \bibinfo{pages}{054301} (\bibinfo{year}{2013}).

\bibitem[{\citenamefont{Zhu et~al.}(2011)\citenamefont{Zhu, Santos, Sankar,
  Chikara, Howard, Chou, Chamon, and El-Batanouny}}]{Zhu:2011ab}
\bibinfo{author}{\bibfnamefont{X.}~\bibnamefont{Zhu}},
  \bibinfo{author}{\bibfnamefont{L.}~\bibnamefont{Santos}},
  \bibinfo{author}{\bibfnamefont{R.}~\bibnamefont{Sankar}},
  \bibinfo{author}{\bibfnamefont{S.}~\bibnamefont{Chikara}},
  \bibinfo{author}{\bibfnamefont{C.~.} \bibnamefont{Howard}},
  \bibinfo{author}{\bibfnamefont{F.~C.} \bibnamefont{Chou}},
  \bibinfo{author}{\bibfnamefont{C.}~\bibnamefont{Chamon}}, \bibnamefont{and}
  \bibinfo{author}{\bibfnamefont{M.}~\bibnamefont{El-Batanouny}},
  \bibinfo{journal}{Phys. Rev. Lett.} \textbf{\bibinfo{volume}{107}},
  \bibinfo{pages}{186102} (\bibinfo{year}{2011}).

\bibitem[{\citenamefont{Sch{\"o}nhense
  et~al.}(2018)\citenamefont{Sch{\"o}nhense, Medjanik, Fedchenko, Chernov,
  Ellguth, Vasilyev, Oelsner, Viefhaus, Kutnyakhov, Wurth
  et~al.}}]{Schonhense:2018aa}
\bibinfo{author}{\bibfnamefont{B.}~\bibnamefont{Sch{\"o}nhense}},
  \bibinfo{author}{\bibfnamefont{K.}~\bibnamefont{Medjanik}},
  \bibinfo{author}{\bibfnamefont{O.}~\bibnamefont{Fedchenko}},
  \bibinfo{author}{\bibfnamefont{S.}~\bibnamefont{Chernov}},
  \bibinfo{author}{\bibfnamefont{M.}~\bibnamefont{Ellguth}},
  \bibinfo{author}{\bibfnamefont{D.}~\bibnamefont{Vasilyev}},
  \bibinfo{author}{\bibfnamefont{A.}~\bibnamefont{Oelsner}},
  \bibinfo{author}{\bibfnamefont{J.}~\bibnamefont{Viefhaus}},
  \bibinfo{author}{\bibfnamefont{D.}~\bibnamefont{Kutnyakhov}},
  \bibinfo{author}{\bibfnamefont{W.}~\bibnamefont{Wurth}},
  \bibnamefont{et~al.}, \bibinfo{journal}{New Journal of Physics}
  \textbf{\bibinfo{volume}{20}}, \bibinfo{pages}{033004}
  (\bibinfo{year}{2018}).

\bibitem[{\citenamefont{Hofmann et~al.}(1996)\citenamefont{Hofmann, Schaff, and
  Schindler}}]{Hofmann:1996aa}
\bibinfo{author}{\bibfnamefont{P.}~\bibnamefont{Hofmann}},
  \bibinfo{author}{\bibfnamefont{O.}~\bibnamefont{Schaff}}, \bibnamefont{and}
  \bibinfo{author}{\bibfnamefont{K.~M.} \bibnamefont{Schindler}},
  \bibinfo{journal}{Physical Review Letters} \textbf{\bibinfo{volume}{76}},
  \bibinfo{pages}{948} (\bibinfo{year}{1996}).

\bibitem[{\citenamefont{Martins et~al.}(2006)\citenamefont{Martins,
  Wellh{\"o}fer, Hoeft, Wurth, Feldhaus, and Follath}}]{Martins:2006aa}
\bibinfo{author}{\bibfnamefont{M.}~\bibnamefont{Martins}},
  \bibinfo{author}{\bibfnamefont{M.}~\bibnamefont{Wellh{\"o}fer}},
  \bibinfo{author}{\bibfnamefont{J.~T.} \bibnamefont{Hoeft}},
  \bibinfo{author}{\bibfnamefont{W.}~\bibnamefont{Wurth}},
  \bibinfo{author}{\bibfnamefont{J.}~\bibnamefont{Feldhaus}}, \bibnamefont{and}
  \bibinfo{author}{\bibfnamefont{R.}~\bibnamefont{Follath}},
  \bibinfo{journal}{Review of Scientific Instruments}
  \textbf{\bibinfo{volume}{77}}, \bibinfo{pages}{115108}
  (\bibinfo{year}{2006}).

\bibitem[{\citenamefont{Sch{\"o}nhense
  et~al.}(2015)\citenamefont{Sch{\"o}nhense, Medjanik, Tusche, {de Loos}, {van
  der Geer}, Scholz, Hieke, Gerken, Kirschner, and Wurth}}]{Schonhense:2015ab}
\bibinfo{author}{\bibfnamefont{G.}~\bibnamefont{Sch{\"o}nhense}},
  \bibinfo{author}{\bibfnamefont{K.}~\bibnamefont{Medjanik}},
  \bibinfo{author}{\bibfnamefont{C.}~\bibnamefont{Tusche}},
  \bibinfo{author}{\bibfnamefont{M.}~\bibnamefont{{de Loos}}},
  \bibinfo{author}{\bibfnamefont{B.}~\bibnamefont{{van der Geer}}},
  \bibinfo{author}{\bibfnamefont{M.}~\bibnamefont{Scholz}},
  \bibinfo{author}{\bibfnamefont{F.}~\bibnamefont{Hieke}},
  \bibinfo{author}{\bibfnamefont{N.}~\bibnamefont{Gerken}},
  \bibinfo{author}{\bibfnamefont{J.}~\bibnamefont{Kirschner}},
  \bibnamefont{and} \bibinfo{author}{\bibfnamefont{W.}~\bibnamefont{Wurth}},
  \bibinfo{journal}{Ultramicroscopy} \textbf{\bibinfo{volume}{159}},
  \bibinfo{pages}{488} (\bibinfo{year}{2015}).

\bibitem[{\citenamefont{Sch{\"o}nhense
  et~al.}(2021)\citenamefont{Sch{\"o}nhense, Kutnyakhov, Pressacco, Heber,
  Wind, Agustsson, Babenkov, Vasilyev, Fedchenko, Chernov
  et~al.}}]{Schonhense:2021ab}
\bibinfo{author}{\bibfnamefont{G.}~\bibnamefont{Sch{\"o}nhense}},
  \bibinfo{author}{\bibfnamefont{D.}~\bibnamefont{Kutnyakhov}},
  \bibinfo{author}{\bibfnamefont{F.}~\bibnamefont{Pressacco}},
  \bibinfo{author}{\bibfnamefont{M.}~\bibnamefont{Heber}},
  \bibinfo{author}{\bibfnamefont{N.}~\bibnamefont{Wind}},
  \bibinfo{author}{\bibfnamefont{S.~Y.} \bibnamefont{Agustsson}},
  \bibinfo{author}{\bibfnamefont{S.}~\bibnamefont{Babenkov}},
  \bibinfo{author}{\bibfnamefont{D.}~\bibnamefont{Vasilyev}},
  \bibinfo{author}{\bibfnamefont{O.}~\bibnamefont{Fedchenko}},
  \bibinfo{author}{\bibfnamefont{S.}~\bibnamefont{Chernov}},
  \bibnamefont{et~al.}, \bibinfo{journal}{Review of Scientific Instruments}
  \textbf{\bibinfo{volume}{92}}, \bibinfo{pages}{053703}
  (\bibinfo{year}{2021}).

\bibitem[{\citenamefont{Redlin et~al.}(2011)\citenamefont{Redlin, Al-Shemmary,
  Azima, Stojanovic, Tavella, Will, and D{\"u}sterer}}]{Redlin:2011aa}
\bibinfo{author}{\bibfnamefont{H.}~\bibnamefont{Redlin}},
  \bibinfo{author}{\bibfnamefont{A.}~\bibnamefont{Al-Shemmary}},
  \bibinfo{author}{\bibfnamefont{A.}~\bibnamefont{Azima}},
  \bibinfo{author}{\bibfnamefont{N.}~\bibnamefont{Stojanovic}},
  \bibinfo{author}{\bibfnamefont{F.}~\bibnamefont{Tavella}},
  \bibinfo{author}{\bibfnamefont{I.}~\bibnamefont{Will}}, \bibnamefont{and}
  \bibinfo{author}{\bibfnamefont{S.}~\bibnamefont{D{\"u}sterer}},
  \bibinfo{journal}{Nuclear Instruments and Methods in Physics Research Section
  A: Accelerators, Spectrometers, Detectors and Associated Equipment}
  \textbf{\bibinfo{volume}{635}}, \bibinfo{pages}{S88} (\bibinfo{year}{2011}),
  ISSN \bibinfo{issn}{0168-9002}, \bibinfo{note}{photonDiag 2010}.

\bibitem[{\citenamefont{Hoffmann et~al.}(2004)\citenamefont{Hoffmann,
  S{\o}ndergaard, Schultz, Li, and Hofmann}}]{Hoffmann:2004aa}
\bibinfo{author}{\bibfnamefont{S.~V.} \bibnamefont{Hoffmann}},
  \bibinfo{author}{\bibfnamefont{C.}~\bibnamefont{S{\o}ndergaard}},
  \bibinfo{author}{\bibfnamefont{C.}~\bibnamefont{Schultz}},
  \bibinfo{author}{\bibfnamefont{Z.}~\bibnamefont{Li}}, \bibnamefont{and}
  \bibinfo{author}{\bibfnamefont{P.}~\bibnamefont{Hofmann}},
  \bibinfo{journal}{Nuclear Instruments and Methods in Physics Research, A}
  \textbf{\bibinfo{volume}{523}}, \bibinfo{pages}{441} (\bibinfo{year}{2004}).

\bibitem[{\citenamefont{Madelung}(2004)}]{Madelung:2004aa}
\bibinfo{author}{\bibfnamefont{O.}~\bibnamefont{Madelung}},
  \emph{\bibinfo{title}{Semiconductors: data handbook}}
  (\bibinfo{publisher}{Springer Science \& Business Media},
  \bibinfo{year}{2004}).

\bibitem[{\citenamefont{Tanuma et~al.}(1994)\citenamefont{Tanuma, Powell, and
  Penn}}]{Tanuma:1994aa}
\bibinfo{author}{\bibfnamefont{S.}~\bibnamefont{Tanuma}},
  \bibinfo{author}{\bibfnamefont{C.~J.} \bibnamefont{Powell}},
  \bibnamefont{and} \bibinfo{author}{\bibfnamefont{D.~R.} \bibnamefont{Penn}},
  \bibinfo{journal}{Surface and Interface Analysis}
  \textbf{\bibinfo{volume}{21}}, \bibinfo{pages}{165} (\bibinfo{year}{1994}).

\end{thebibliography}

\end{document}